\documentclass[10pt]{iopart}
\usepackage{graphicx}
\usepackage{amssymb}
\eqnobysec

\newcommand{\beq}{\begin{equation}}
\newcommand{\eeq}{\end{equation}}
\newcommand{\be}{\begin{equation*}}
\newcommand{\ee}{\end{equation*}}
\newcommand{\beqa}{\begin{eqnarray}}
\newcommand{\eeqa}{\end{eqnarray}}
\newcommand{\bea}{\begin{eqnarray*}}
\newcommand{\eea}{\end{eqnarray*}}

\usepackage[colorlinks,linkcolor=blue,urlcolor=black,citecolor=blue]{hyperref}

\def\stackunder#1#2{\mathrel{\mathop{#1}\limits_{#2}}}
\newcommand{\stackover}[2]{\mathrel{\mathop{#1}^{#2}}}

\newcommand{\abs}[1]{\vert#1\vert}
\newcommand{\bigac}[1]{\left(#1\right)}
\newcommand{\bigmean}[1]{\left\langle#1\right\rangle}
\newcommand{\dd}{{\rm d}}
\newcommand{\eps}{{\varepsilon}}
\newcommand{\erf}{\mathop{\rm erf}}
\newcommand{\half}{{\scriptstyle{\scriptstyle1\over\scriptstyle2}}}
\newcommand{\ii}{{\rm i}}
\newcommand{\lap}[1]{\mathrel{\mathop{\cal L}\limits_{#1}^{}}}
\renewcommand{\max}{{\rm max}}
\newcommand{\mean}[1]{\langle#1\rangle}
\newcommand{\prob}{\mathbb{P}}
\renewcommand{\th}{{\theta}}
\newcommand{\var}{\mathop{\rm Var}\nolimits}
\newcommand{\w}[1]{{\widetilde{#1}}}

\newcommand{\tone}{{\bf t}_1}
\newcommand{\vone}{v_{\tone}\!}
\newcommand{\rr}{r}

\begin{document}

\title{Record statistics of integrated random walks and the random acceleration process}

\author{Claude Godr\`eche and Jean-Marc Luck}

\address{Universit\'e Paris-Saclay, CNRS, CEA, Institut de Physique Th\'eorique,
91191~Gif-sur-Yvette, France}

\begin{abstract}

We address the theory of records for integrated random walks with finite variance.
The long-time continuum limit of these walks is a non-Markov process known as the random acceleration process or the integral of Brownian motion.
In this limit, the renewal structure of the record process
is the cornerstone for the analysis of its statistics.
We thus obtain the analytical expressions of several characteristics of the process, notably the distribution of the total duration of record runs (sequences of consecutive records), which is the continuum analogue of the number of records of the integrated random walks.
This result is universal, i.e., independent of the details of the parent distribution of the step lengths.

\end{abstract}

\eads{\mailto{claude.godreche@ipht.fr},\mailto{jean-marc.luck@ipht.fr}}

\maketitle

\section{Introduction}
\label{intro}

An observation in a time series is called an (upper) record if it is greater
than all previous observations in the series.
Two simple situations arise either when the observations are independent
identically distributed (iid) random variables $\eta_1,\eta_2,\dots,\eta_n$,
drawn from a given continuous distribution $\Phi$,
or when they are the successive positions of a random walk
$V_1,V_2,\dots,V_n$ built from the previous variables, with
\be
V_{n}-V_{n-1}=\eta_n,
\ee
hence
\beq
V_n=V_0+\eta_1+\cdots+\eta_n,
\label{vedef}
\eeq
where the number $n$ of steps is a discrete time.
The number $M_n$ of records scales as $\ln n$ in the first case, 
independently of the choice of the step length distribution $\Phi$,
and as $\sqrt{n}$ in the second case,
again independently of this distribution,
if, e.g., the latter is symmetric.\footnote{Further details
and a comprehensive bibliography are
presented a little further down in this introduction.}

Consider now the \textit{integrated random walk} $X_1,X_2,\dots,X_n$ defined as
\be
X_n-X_{n-1}=V_n,
\ee
or
\be
X_{n}-2X_{n-1}+X_{n-2}=\eta_n,
\ee
hence
\beq
X_n=X_0+V_1+\cdots+V_n.
\label{xdef}
\eeq
While the $V_n$ process is Markovian, the $X_n$ process does not possess this property.
Nonetheless the couple $(V_n,X_n)$ is Markovian, since at each step its
evolution is entirely determined by the noise $\eta_n$, as the following
recursion shows
\be
(V_0,X_0)\stackover{\longrightarrow}{\eta_1}(V_1,X_1)\stackover{\longrightarrow}{\eta_2}(V_2,X_2)\stackover{\longrightarrow}{\eta_3}\cdots
\ee

\begin{figure}[!ht]
\begin{center}
\includegraphics[angle=0,width=.8\linewidth,clip=true]{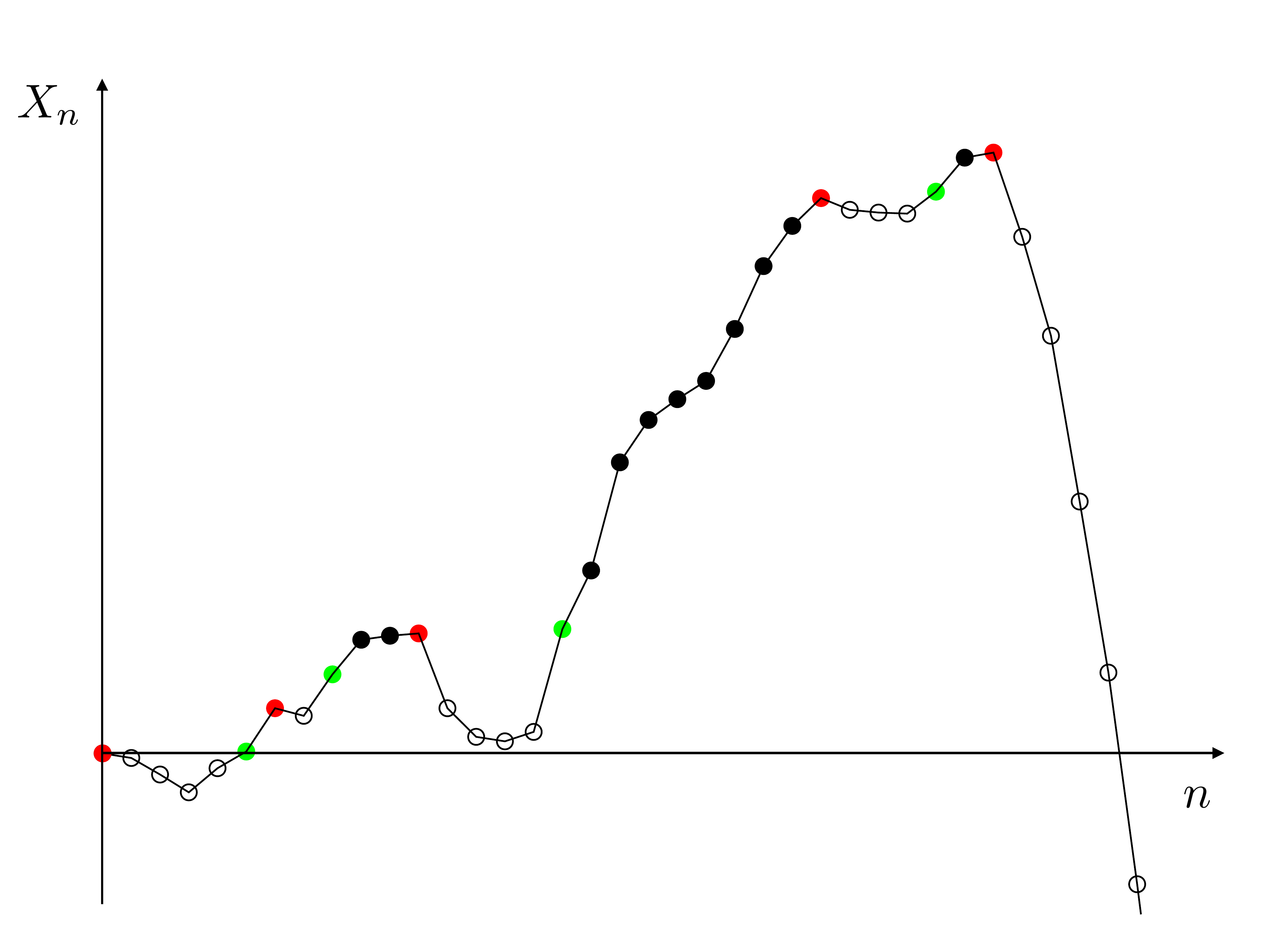}
\caption{\small
A sample path of an integrated random walk $X_n$ defined according to (\ref{vedef}), (\ref{xdef}) with a symmetric Gaussian distribution of steps, launched from $(V_0,X_0)=(0,0)$, up to $n=36$.
The full symbols are the successive records of the walk $X_n$.
Record runs (sequences of consecutive records) begin at a green dot and end at a red one.
The number $M_{n}$ of records in this example is equal to $19$.
}
\label{fig:figure0-X}
\end{center}
\end{figure}
\begin{figure}[!ht]
\begin{center}
\includegraphics[angle=0,width=.8\linewidth,clip=true]{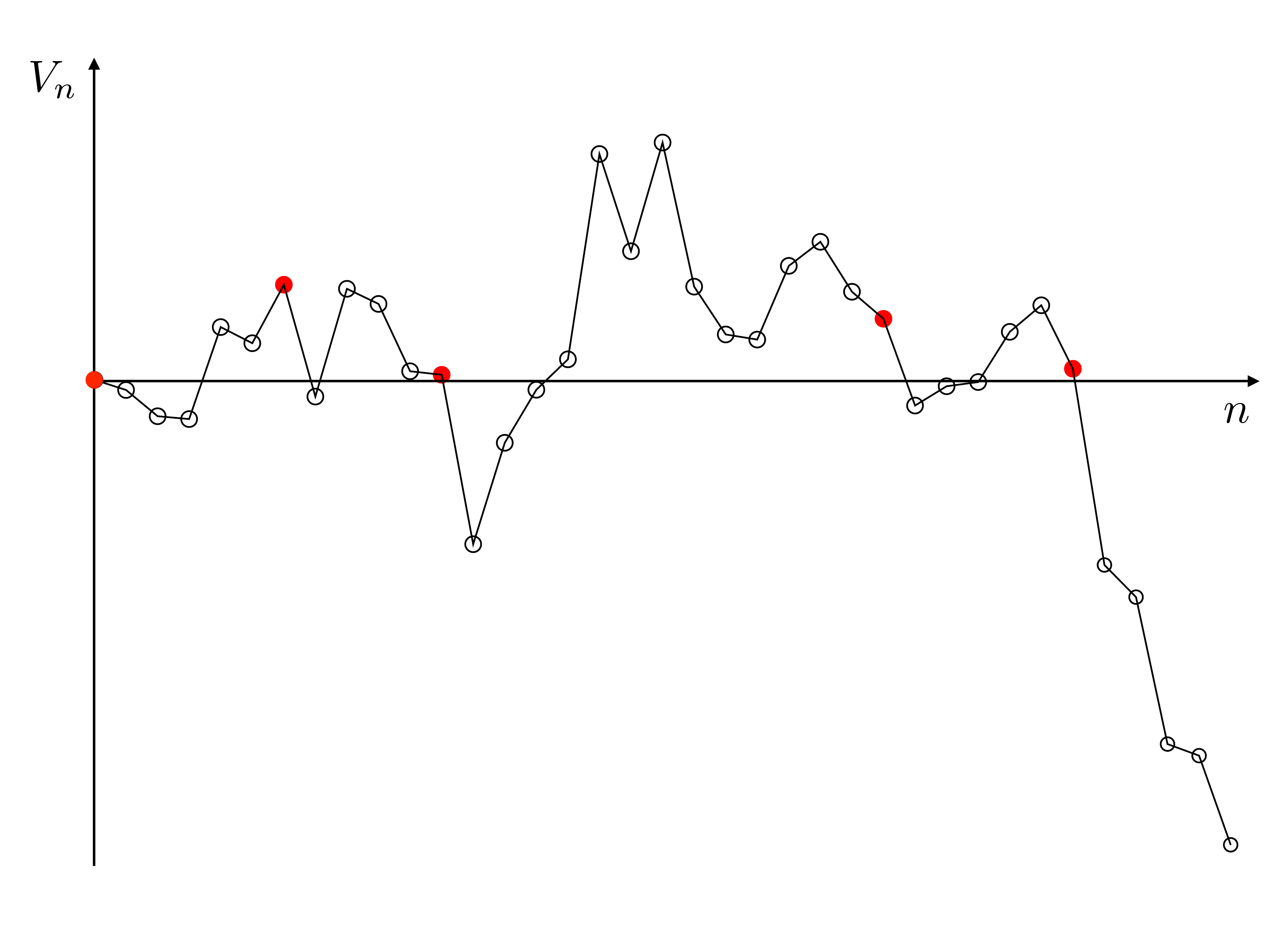}
\caption{\small
Corresponding path of the random walk $V_n$.
Red dots are marked at the same epochs as the red dots of figure \ref{fig:figure0-X}.
They correspond to the left endpoints of downcrossing steps, except maybe at the origin.
}
\label{fig:figure0-V}
\end{center}
\end{figure}

The aim of the present work is to investigate the statistics of records for
the integrated random walk $X_n$.
We assume henceforward that the step length distribution $\Phi$ is symmetric
with finite variance $\mean{\eta^2}=2D$, i.e., such that the random walk $V_n$ is diffusive, with diffusion coefficient $D$.
For this class of distributions, 
asymptotic properties of the random walk $V_n$
and of the integrated random walk $X_n$
are described by their continuum analogues,
Brownian motion, and the integral of Brownian motion---or random acceleration process---respectively.
As will be seen below, the number $M_n$ of records for the integrated process scales as $n$.
We shall focus our attention on the exact asymptotic distribution of this number of records,
and of related observables,
in the limit of long times.

A sample path of such an integrated random walk $X_n$ is depicted in figure \ref{fig:figure0-X} and the
corresponding path of the random walk $V_n$ is depicted in figure \ref{fig:figure0-V}.
These figures reveal some salient features.
In figure \ref{fig:figure0-X} full symbols are the successive records of $X_n$.
Record runs (sequences of consecutive records) begin at a green dot and end at a red one.
In figure \ref{fig:figure0-V} red dots are marked at the same epochs as the red dots of figure \ref{fig:figure0-X}.
They correspond to the beginnings of downcrossing steps.
Thus, at red dots the process $(V_n,X_n)$ \textit{almost} restarts afresh.
This renewal structure is indeed only approximate, because at red dots the positions of the walker $V_n$ are positive instead of being exactly zero, as at the starting point $V_0$.
However, as we shall see, in the continuum limit,
sections between two red dots become probabilistic replicas of each other.
This observation will constitute the basis of the analysis to come.

Before proceeding, we take a step back and give a
brief account of the subject in order to put this work into context.

\subsection*{A brief survey of the statistics of records}

As recalled above, an observation in a time series is called an upper (lower)
record if it is greater (smaller) than all previous observations in the series.
The first study of this topic, for the case of independent observations---that
its to say, of iid random variables---is due to
Chandler~\cite{chandler}.
His study triggered a number of subsequent works on records for
random observations, either for statistical purposes or aimed at more
theoretical probabilistic investigations.
Independently, R\'enyi, in~\cite{renyi}, laid the foundations of a more
theoretical approach to the statistics of records for iid random variables.
The study of records in such a situation grew into a large body of knowledge
now referred to as the \textit{classical theory of records}~\cite{chandler,renyi,foster,
glick,arnold,nevzorov2,bunge,nevzorov}.
The last of these references gives an account of the literature on the subject
at the turn of the century.

A basic knowledge of this subject is easy to grasp (see the references above).
Consider again the sequence of iid random variables, $\eta_1,\eta_2,\dots$.
By definition, the variable $\eta_n$ is a record if
\be
\eta_n>\max(\eta_1,\dots,\eta_{n-1}),
\ee
and the label $n$ is referred to as a record time.
The first value $\eta_1$ is considered as a record.
Since this definition only involves inequalities between
the variables~$\eta_n$, the statistics of record times is independent of
the underlying distribution, provided that it is continuous.
In particular, the occurrence of a record at any time $n\ge2$ has probability 
\be
\rr_n=\frac{1}{n}.
\ee
Indeed, amongst the $n!$ permutations of $\eta_1,\dots,\eta_n$,
there are $(n-1)!$ permutations where $\eta_n$ is the largest.
In terms of the indicator variables $I_n$,
equal to 1 if $\eta_n$ is a record and to 0 otherwise,
the number $M_n$ of records up to time $n$
reads
\beq\label{eq:MnI}
M_n=I_1+I_2+\cdots+I_n,
\eeq
with $\rr_n=\mean{I_n}$.
Thus\footnote{See \ref {app:word} for the notations used in the present work.}
\be
\mean{M_n}
=\sum_{i=1}^n \rr_i
=\sum_{i=1}^n\frac{1}{i}
\stackunder{\approx}{n\to\infty}\ln n+\gamma,
\ee
where $\gamma=0.577\,215\dots$ is Euler's constant.
The variance of $M_n$ also scales as $\ln n$, with unit prefactor,
as well as all higher order cumulants.
The bulk of the distribution of $M_n$ has the asymptotic Poissonian form 
\beq
\prob(M_n=m)\stackunder{\approx}{n\to\infty}\frac{1}{n}\frac{(\ln
n)^{m-1}}{(m-1)!},
\label{amn}
\eeq
confirming the $\ln n$ scaling mentioned earlier.

Records for one-dimensional random walks constitute a second facet of the
theory of records,
whose developments ran parallel to and independently from the studies of
records for iid random variables.
A remarkable historical coincidence, which seemingly has gone unnoticed so far,
is that the paper of Chandler~\cite{chandler} on records for iid random
variables and the paper by Blackwell~\cite{blackwell}, which laid the
foundations of the study of records for random walks, were both simultaneously
presented for publication in June 1952.
Blackwell introduced the times of occurrence and values of records,
`\textit{aptly christened the ladder random variables of the random walk}' by
Feller in~\cite{feller1}, to quote Spitzer~\cite{spitzer}.

As a matter of fact, looking at the occurrences of `\textit{record values}' in
the monographs of Feller~\cite{feller1,feller2}, it is found that this
terminology refers, on the one hand, as expected, to the simple situation of
iid random variables described above, but also, on the other hand, to the
\textit{ladder points} of one-dimensional random walks.
To quote Feller~\cite[Ch.~XII]{feller2}, `\textit{Looking at the graph of a
random walk one notices as a striking feature the points where $S_n$} [the
position of the walk after $n$ steps, denoted as $V_n$ above] \textit{reaches a
record value, that is, where $S_n$ exceeds all previously attained values $S_0,\dots,S_{n-1}$.
These are the ladder points (...).
The theoretical importance of ladder points derives from the fact that the
sections between them are probabilistic replicas of each other, and therefore
important conclusions concerning the random walk can be derived from a study of
the first ladder point}'.
As emphasised by Spitzer~\cite{spitzer}, it was in connection with renewal
theory that ladder random variables were first studied by Blackwell.
Renewal theory~\cite{feller1,feller2,cox,cox-miller} is precisely the
cornerstone for the investigation of the statistics of
records---or ladder points---for random walks, as the second sentence of
Feller quoted above suggests, and as is amply detailed in chapters XII and
XVIII of~\cite{feller2}.

All the tools necessary to investigate the statistics of records for random walks are contained in~\cite{feller2}.
In a nutshell, the distribution $f(n)$ of waiting times between two records is given by Sparre Andersen theory~\cite{sparre53,sparre54,feller2}.
This distribution is universal, i.e., independent of the parent distribution $\Phi$, provided the latter is continuous and symmetric. 
Hence the distribution of the number of records $M_n$ after $n$ steps of the walk (or in other words the number of renewals up to time $n$), is universal, too.
The expression of its generating function 
\beq\label{eq:genfctn}
\sum_{n\ge0} \prob(M_n=m)\, z^n = {\tilde f}(z)^{m-1}\, 
\frac{1-\tilde f(z)}{1-z},
\eeq
(where $\tilde f(z)=1-\sqrt{1-z}$ is the generating function of the $f(n)$)
is an immediate consequence of the renewal structure of the sequence of records (or ladder points) (see, e.g.,~\cite[\S~3]{glrenew},~\cite{ziff},~\cite[\S~3]{revue}).
From (\ref{eq:genfctn}) the mean number of records ensues easily,
\be
\mean{M_n} \stackunder{\approx}{n\to\infty} \frac{2\sqrt{n}}{\sqrt{\pi}},
\ee
as well as the expression of $\prob(M_n=m)$, whose asymptotic scaling form reads
\be
\prob(M_n=m)\stackunder{\approx}{n\to\infty} \frac{1}{\sqrt{n}}\,g\left(\frac{m}{\sqrt{n}}\right) , \qquad
 g(x)= \frac{\e^{-x^2/4}}{\sqrt{\pi}}\; (x\ge0),
\ee
(see, e.g.,~\cite[page 373]{feller2},~\cite[\S~3]{glrenew},~\cite{ziff},~\cite[\S~3]{revue}), confirming the $\sqrt{n}$ scaling mentioned earlier.
The number of records can still be expressed as the sum (\ref{eq:MnI}), however now the indicator variables $I_1,I_2,\dots$ are no longer independent.
The probability $r_n$ of occurrence of a record at time $n$ is still equal to $\mean{I_n}$, with the explicit expression~\cite{revue}
\be
\rr_n=q(n)=\frac{1}{2^{2n}} {2n \choose n} \stackunder{\approx}{n\to\infty}\frac{1}{\sqrt{\pi\,n}},
\ee
where $q(n)$, the probability that the random walk, starting at the 
initial position $V_0$, stays below $V_0$ up to step $n$, is related to the first passage probability $f(n)$ by
$f(n)= q(n-1)-q(n)$.

More recently, investigations on the theory of records have permeated the physics community
as being part of the broader field of extremal events, of natural interest in physics.
Records can indeed be seen as extremal events monitored in time.
We refer the reader to~\cite{revue,wergen} for an overview
of the recent applications of the theory of records in statistical physics.
In particular the theory of records for one-dimensional random walks has been
revisited and enriched in a series of papers in the past two decades (see~\cite{revue} and the references therein).

In contrast, to our knowledge, essentially nothing is known so far on the statistics of records for the integrated
random walk or for the random acceleration process.
These processes are known to be appreciably more difficult to study than random walks.
A number of advances have nevertheless been made in the past or more recently
on various aspects of these processes, both in mathematics~\cite{mckean,goldman,marshall,lachal,sinai,profeta} and in physics, in particular on first-passage
properties, inelastic collapse of particles, time at which the maximum is reached, statistics of the occupation time, 
dynamics with resetting~\cite{burk1,bray,burk2,smedt,rosso,burk3,singh}
(see the recent review~\cite{burk4} for additional references).

The question of analysing the statistics of records for the integrated random
walk was raised in~\cite{revue} but left unanswered. 
It is the purpose of the present work to fill this gap, at least for diffusive random walks,
where the variance of the step lengths is finite.
We start with a summary of our findings.

\subsection*{Summary of the results of the present work}

\begin{figure}[!ht]
\begin{center}
\includegraphics[angle=0,width=.9\linewidth,clip=true]{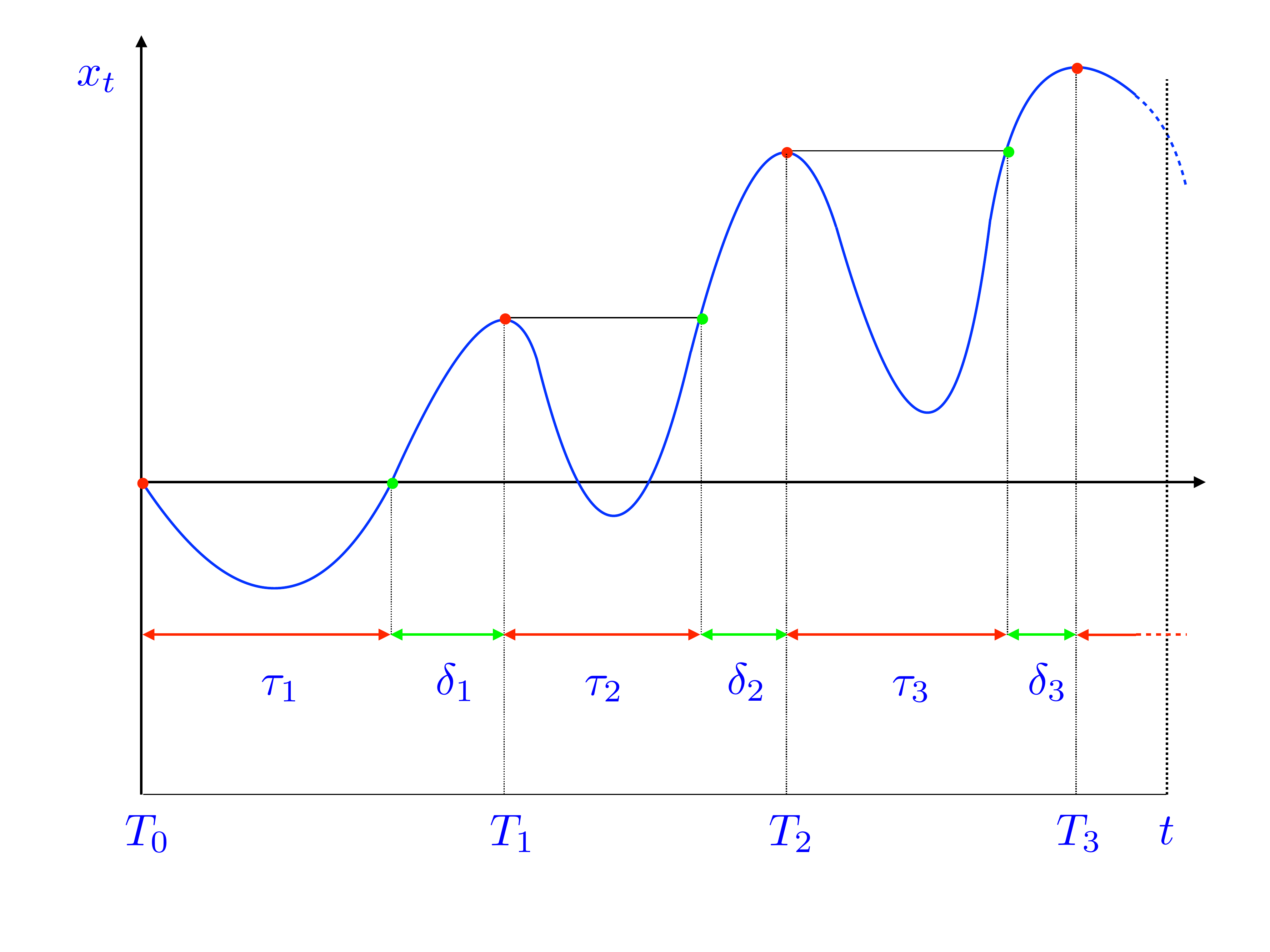}
\caption{\small
Schematic drawing of a path of the random acceleration process.
Sections between two red dots are probabilistic replicas of each other.
In each section, records for $x_t$ take place between a green and a red 
dot---this defines the lengths $\delta_n$ of the record runs.
In the present example, discarding the red dot at the origin, the number $N_t$ of renewals (red dots) between 0 and $t$ is equal to $3$.
}
\label{fig:renew}
\end{center}
\end{figure}

We shall show that for the random acceleration process, which is the continuum analogue of the integrated random walks defined above,
the process of records has a renewal structure
involving an infinite sequence of iid couples of intervals of time $(\tau_n,\delta_n)$,
whose joint law is given in (\ref{eq:brique}).
As depicted in figure \ref{fig:renew}, the endpoints of the time intervals $\tau_n$ (respectively,~$\delta_n$)
are marked by green dots (respectively, red dots).
A record run (a sequence of consecutive records) starts at every green dot
and stops at the next red one.
This renewal structure---foreshadowed by the discrete process---is the cornerstone for the analysis that follows.
Henceforth we shall call the red dots renewal events (or renewals for short).

The simplest observable to consider is the number $N_t$ of renewals, or equivalently of records runs, occurring between 0 and $t$.
The determination of its distribution is a classic in renewal theory (see, e.g.,~\cite{feller2,glrenew}).
In particular its average grows as
\be
\mean{N_t}\approx A\,t^{1/4},
\ee
where the exponent 1/4 is the well-known persistence exponent
of the random acceleration process~\cite{goldman,sinai,burk1},
whereas the non-universal prefactor $A$ depends on the parent distribution 
$\Phi$, as demonstrated in Table \ref{tab:aeps}.

As previously stated, $M_n$, the number of records up to time $n$ in the discrete theory, grows linearly with $n$.
More precisely,
its distribution scales as
\beq
\prob(M_n=m)\stackunder{\approx}{n\to\infty}
\frac{1}{n}\,f_R\left(\frac{m}{n}\right),
\eeq
where the universal scaling function $f_R$ is the probability density of the ratio
\beq\label{eq:rratio}
R=\lim_{n\to\infty}\frac{M_n}{n}=\lim_{t\to\infty}\frac{S_t}{t},
\eeq
in which $S_t$, the sum of all (finished or not finished) record runs, is the continuum analogue of $M_n$.
Thus $R$ is the fraction of time during which the process sets a record.
The density $f_R(x)$ is an asymmetric U-shaped curve depicted in figure \ref{fig:frhisto}, whose analytic expression is given in (\ref{eq:fR}).
Its first moment 
\beq\label{eq:meanR}
\mean{R}=\rr_\infty
=1-\frac{\sqrt{6}}{4}=0.387\,627\dots
\eeq
is a universal number, whose natural interpretation is the limit, when the discrete time $n\to\infty$, of the
probability $\rr_n$ of occurrence of a record at $n$, or probability of record breaking, defined as for the cases of iid random variables or random walks,
\beq\label{eq:defQn}
\rr_n=\prob(X_n>\max(X_1,\dots,X_{n-1})).
\eeq
In other words,
\beq\label{eq:Qinfty}
\mean{R}=\lim_{n\to\infty}\frac{\mean{M_n}}{n}=\lim_{n\to\infty}\frac{1}{n}\sum_{i= 1}^n \rr_i
=\rr_\infty.
\eeq
As a consequence, the mean number of records grows linearly as
\be
\mean{M_n}\stackunder{\approx}{n\to\infty} n\,\rr_\infty.
\ee
We complement this study by the determination of the distributions of the epochs $T_{N_t}$, of the last renewal before $t$,
and $\w T_{N_t}$, of the last dot before $t$, regardless of its colour.
Both observables again grow linearly with time.
The densities $f_U(x)$ and $f_V(x)$ of the corresponding limiting ratios
\be
U=\lim_{t\to\infty}\frac{T_{N_t}}{t},\qquad V=\lim_{t\to\infty}\frac{\w T_{N_t}}{t},
\ee
 given in (\ref{eq:fU})
and (\ref{eq:fV}) and depicted in figure \ref{fig:fehisto}, are universal.
Armed with this knowledge
we recover, by a different method, a result given in~\cite{rosso} on the time $t_{\rm m}$ for the random acceleration process to reach its maximum.
The density of the rescaled time
\be
W=\lim_{t\to\infty}\frac{t_{\rm m}}{t}
\ee
has a universal expression given in (\ref{eq:tm}), involving $\rr_\infty$ and the density $f_U$, which sheds new light on the result given in~\cite{rosso}.

A last remark is in order.
In the present study, universality for integrated random walks only manifests itself asymptotically.
This is particularly evident when considering the probability of record breaking $\rr_n$, defined in (\ref{eq:defQn}).
Except for 
$\rr_1=\prob(X_1>0)=1/2$,
the probability $\rr_n$ of having a record at any finite $n$ is non universal.
For instance, for $n=2$, $X_2=V_1+V_2=2\eta_1+\eta_2$, thus, using symmetries, we have
\beq\label{q2res}
\rr_2=\frac12-\prob(\eta_1>0, \eta_1<\eta_2<2\eta_1).
\eeq
This probability is non-universal, as demonstrated in
Table \ref{q2}.
Universality is reached asymptotically, i.e., $\rr_n\to\rr_\infty$.

\begin{table}[!ht]
\begin{center}
\begin{tabular}{|l|l|}
\hline
distribution & $\rr_2$ \cr
\hline
uniform & $7/16=0.4375$ \cr
triangular & $43/96=0.447\,916\,\dots$ \cr
exponential & $11/24=0.458\,333\,\dots$ \cr
Gaussian & $5/8-(\arctan 2)/(2\pi)=0.448\,791\,\dots$ \cr
\hline
binary & $1/4=0.25$ \cr
\hline
\end{tabular}
\caption
{Exact value of the probability $\rr_2$ of having a record at time $n=2$
for various symmetric step length distributions $\Phi$:
uniform, triangular (the law of the sum of two uniform variables),
exponential, Gaussian and binary $(\pm1)$.
The latter distribution, besides the fact that it is not continuous, hence that (\ref{q2res}) does not hold,
appears as an outlier,
inasmuch as the value $\rr_2=\prob(\eta_1=\eta_2=+1)=1/4$
 is quite different from
those for the continuous distributions, which vary over a rather narrow range.
}
\label{q2}
\end{center}
\end{table}

The paper is structured as follows.
Section \ref{sec:remind} gives preliminaries on the random acceleration process, with emphasis on the distributions of three important random variables, depicted in figure \ref{fig:figure-def}.
Section \ref{sec:renewal} highlights the renewal structure underlying the process of records in the random acceleration process.
Section \ref{sec:nren} provides an analytical treatment of the distribution of the number of renewals.
Section \ref{sec:nrec}, which is the main section, gives the exact distribution of the number of records in the asymptotic regime.
Section \ref{sec:nepoch} gives the distributions of the epochs $T_{N_t}$ and $\w T_{N_t}$ and the distribution of the time at which the random acceleration process reaches its maximum.
We discuss some possible extensions in section \ref{discussion}.
Three appendices contain more technical matters.

\section{Elements on the random acceleration process}
\label{sec:remind}

Throughout this work,
asymptotic analytical results on the statistics of records of integrated random walks
with finite variance
will be obtained using their continuum limit,
the random acceleration process.

\subsection{Definition of the process}

Consider a particle submitted to a random force,
whose position~$x_t$ obeys the stochastic equation of motion
\beq
\frac{\dd^{2}x_{t}}{\dd t^{2}}=\eta_{t},
\label{def1}
\eeq
where $\eta_{t}$ is a normalised Gaussian white noise, i.e.,
\be
\left\langle\eta_{t}\right\rangle =0,\qquad\left\langle\eta_{t}\eta
_{t^{\prime}}\right\rangle =\delta(t-t^{\prime}).
\ee
This is the original Langevin equation without damping force.
Equivalently, the position $x_t$ of the particle and its velocity $v_t$
jointly obey the first-order (i.e., Markovian) dynamics
\beq\label{def2}
\frac{\dd v_{t}}{\dd t}=\eta_{t},\qquad\frac{\dd x_{t}}{\dd t}=v_{t},
\eeq
with initial condition ($v_{0},x_{0}$).
Hence
\be
v_{t}=v_{0}+W_{t},\qquad x_{t}=x_{0}+v_{0}t+\int_{0}^{t}\dd u\,W_{u},
\ee
where the integral of the noise,
\be
W_{t}=\int_{0}^{t}\dd u\,\eta_{u}
\ee
is normalised Brownian motion, such that $D=1/2$.
The process $x_{t}$ is usually referred to as the \textit{integral of Brownian motion} or
the \textit{random acceleration process}.
We have
\be
\mean{v_t}=v_0,\quad
\mean{x_t}=x_0+v_0t,\quad
\var v_t=\mean{W_t^2}=t,\quad
\var x_t=\frac{t^3}{3}.
\ee
The fluctuating parts are the leading ones at long times,
so that $v_t$ and $x_t$ respectively grow as $t^{1/2}$ and as $t^{3/2}$.

\subsection{Some preliminary results}

\begin{figure}[!ht]
\begin{center}
\includegraphics[angle=0,width=.9\linewidth,clip=true]{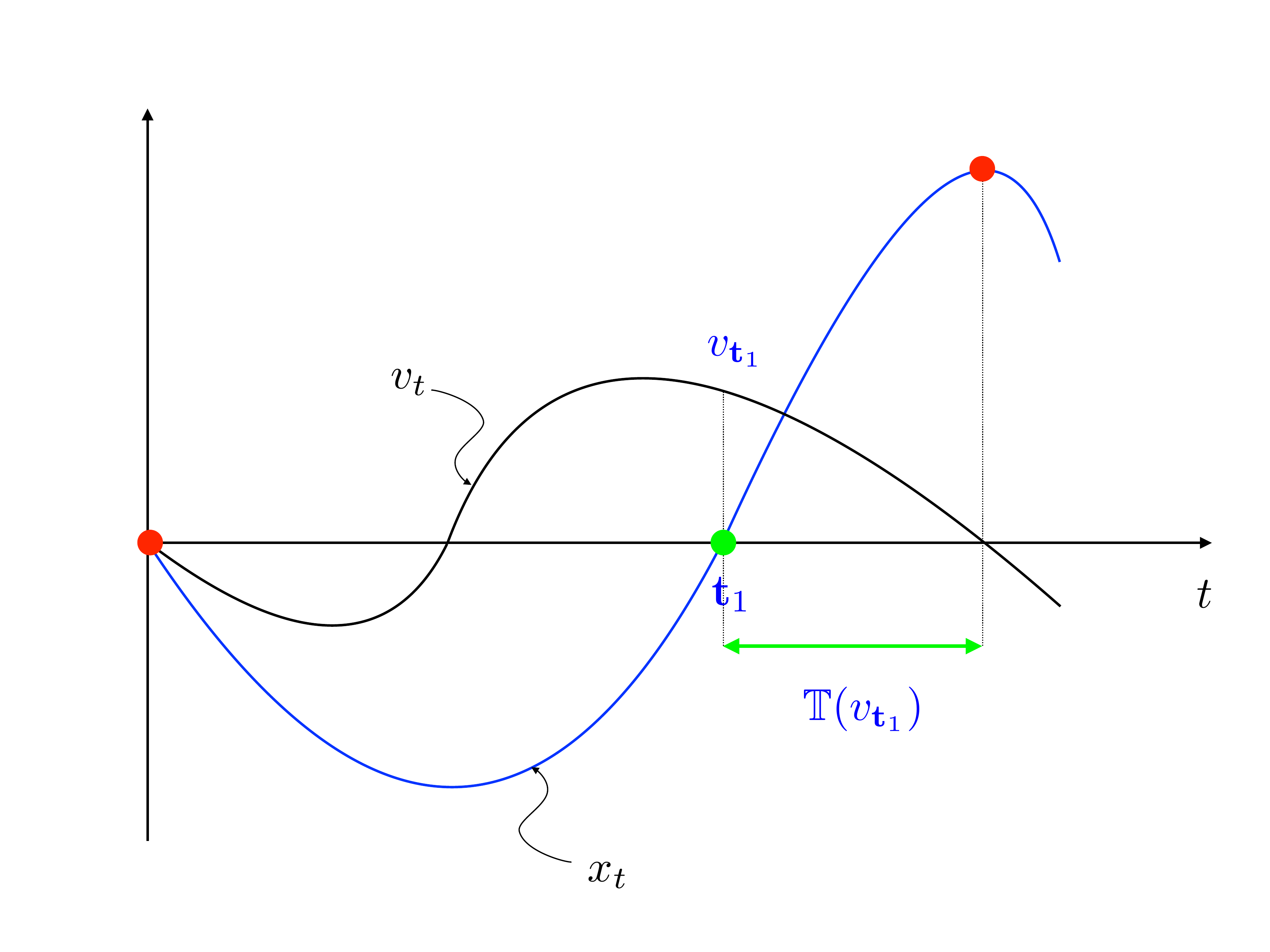}
\caption{\small
Three fundamental random variables.
Epoch of the first zero crossing $\tone$ of $x_t$;
corresponding velocity $\vone$ of the randomly accelerated particle (or position of Brownian motion);
first hitting time $\mathbb{T}(\vone)$ of the origin for Brownian motion starting from $\vone$ at time $t=0$.
(Schematic drawing.)
}
\label{fig:figure-def}
\end{center}
\end{figure}

We gather here some results that we shall need in the sequel.
Consider the randomly accelerated particle starting from the origin with initial velocity $v_0$.
From now on, $v_0$ will represent an initial microscopic velocity acting as a cutoff, the role of which is to regularise the theory, as is necessary when dealing, for example, with first-passage observables in Brownian motion (see section \ref{sec:cutoff} for a detailed discussion).
We shall henceforth use the notation
\beq\label{epsdef}
\eps=\sqrt{|v_0|}.
\eeq
We shall also restrict ourselves to paths of $x_t$ with $v_0$ negative, as in figure \ref{fig:figure-def}, which, as explained later, is a simplification for the analysis which follows.

As will be made clear in the next section, there are three fundamental random variables to consider for the sequel (see figure \ref{fig:figure-def}).
\begin{description}
\item[---] The first one is the time of occurrence of the first zero crossing $\tone$ of $x_t$.
\item[---] The second one is the corresponding velocity $\vone$ of the randomly accelerated particle (or position of Brownian motion).
\item[---] 
The third one is the first hitting time 
$\mathbb{T}(\vone)$ of the origin for Brownian motion starting from $\vone$ at time $t=0$.
Note that this random variable depends itself on another random variable (namely $\vone$).
This is also the time taken by $x_t$ to go from the green dot to the red one, starting from $x_{\tone}=0$ (see figure \ref{fig:figure-def}).
\end{description}

The expression of the joint distribution of the epoch $\tone$ and of the corresponding velocity $\vone$
is a classical result due to McKean~\cite{mckean,lachal} which states that, taking $v_0<0$ and $v>0$ as in figure \ref{fig:figure-def},
\beq\label{eq:jointe}
f_{\tone,\vone}(\tau,v)=\frac{\sqrt{3}\,v}{\pi\tau^2}
\e^{-2(v^2-|v_0|v+v_0^2)/\tau}\erf\sqrt{\frac{6\abs{v_0}v}{\tau}}.
\eeq
The marginal distribution of $\vone$ ensues by integration of (\ref{eq:jointe}) on $\tau$:
\beq\label{eq:margin}
f_{\vone}(v)=\frac{3\,\eps}{2\pi}\frac{v^{3/2}}{v^3+|v_0|^3}.
\eeq
There is no explicit expression of the marginal distribution of $\tone$, henceforth denoted for short as
\be
\rho(\tau)\equiv f_{\tone}(\tau),
\ee
however, in the regime where both $\tau$ and $v$ are large, such that $|v_0|\ll\tau\sim v^2$,
 (\ref{eq:jointe}) simplifies to
\beq\label{eq:ftauv}
f_{\tone ,\vone }(\tau,v)
\approx\frac{6\sqrt{2}\,\eps}{\pi^{3/2}}\,\frac{v^{3/2}}{\tau^{5/2}}\,\e^{-2v^2/\tau}.
\eeq
Thus, by integration on $v$, the asymptotic expression of the marginal $\rho(\tau)$ ensues
\beq\label{eq:rhotau}
\rho(\tau)\stackunder{\approx}{\tau\to\infty} \frac{c}{\tau^{5/4}},
\qquad c=\frac{3\,\Gamma(1/4)\,\eps}{2^{11/4}\pi^{3/2}}.
\eeq

As for the third variable, we have
the well-known result for the probability density of the first hitting time
$\mathbb{T}(v)$ of the origin for Brownian motion starting from $v>0$~\cite{feller2},
\beq\label{eq:hitting0}
f_{\mathbb{T}(v)}({\delta})
=\frac{v\,\e^{-{v^2}{/(2{\delta}})}}{\sqrt{2\pi{\delta}^3}},
\eeq
which thereby provides the expression of 
the conditional probability density of the first hitting time
$\mathbb{T}(\vone)$ of the origin for Brownian motion starting from $\vone=v$,
\beq\label{eq:hitting}
f_{\mathbb{T}(\vone)|\vone}({\delta|v})
=\frac{v\,\e^{-{v^2}{/(2{\delta}})}}{\sqrt{2\pi{\delta}^3}}.
\eeq
Using (\ref{eq:margin}) and (\ref{eq:hitting}), the density $f_{\mathbb{T}(\vone)}(\delta)$ is obtained in the form of the integral of the product
\beq\label{eq:taildelta}
f_{\mathbb{T}(\vone)}(\delta) =\int\dd v\,f_{\mathbb{T}(\vone)|\vone}({\delta|v})\, f_{\vone}(v)
=\frac{1}{\delta}\,g\!\bigac{\frac{\eps}{\delta^{1/4}}},
\eeq
which leads to an explicit albeit complicated expression of $g(x)$.
For $x\to0$ this function behaves as
\be
g(x)\stackunder{\approx}{x\to0}\frac{3\Gamma(1/4)\,}{2^{9/4}\pi^{3/2}}\, x,
\ee
which implies that
\beq\label{eq:fdelta}
f_{\mathbb{T}(\vone)}(\delta) 
\stackunder{\approx}{\delta\to\infty}\frac{c\sqrt{2}}{\delta^{5/4}}.
\eeq
Note that while the tail index of the conditional density (\ref{eq:hitting}) (where $\vone=v$ has a fixed value) is equal to $1/2$, the tail index of the density (\ref{eq:taildelta}) of the random variable $\mathbb{T}(\vone)$ is equal to $1/4$. 

Anticipating on what follows, we shall be interested in the joint density of $\tone$ and $\mathbb{T}(\vone)$.
This density, denoted for short by $\rho(\tau,\delta)$, is obtained by integration of the product of (\ref{eq:jointe}) and (\ref{eq:hitting}):
\beq\label{eq:mult}
\rho(\tau,\delta)\equiv
f_{\tone ,\mathbb{T}(\vone )}(\tau,\delta)=
\int\dd v\,
f_{\mathbb{T}(\vone)|\vone}({\delta|v}) \,
f_{\tone ,\vone}(\tau,v).
\eeq
Its asymptotic expression can be obtained by multiplying (\ref{eq:ftauv}) by (\ref{eq:hitting}) and integrating upon $v$, 
which leads to the scaling form 
\beq\label{eq:brique}
\rho(\tau,\delta)\approx
\frac{9\,\Gamma(3/4)\,\eps}{2^{1/4}\pi^2}\,
\frac{\delta^{1/4}}{\tau^{3/4}(\tau+4\delta)^{7/4}}.
\eeq
This expression is a key ingredient in all subsequent developments.

 The marginals ensuing from (\ref{eq:brique}) by integration upon each variable yield (\ref{eq:rhotau}) and (\ref{eq:fdelta}) back.
The intervals of time $\tone$ and $\mathbb{T}(\vone)$ separately have heavy-tailed distributions
with tail coefficients proportional to $\eps$ and the same tail index $1/4$.
The latter is the well-known survival (or persistence) exponent 
of the random acceleration process~\cite{mckean,goldman,sinai,burk1}, i.e., the decay exponent 
of the survival probability, or probability that the process $x_t$ has not returned to its starting point $x=0$
until time $t$, 
\beq\label{eq:survie}
p(t)=\prob(\tone>t)=\int_{t}^\infty\dd\tau\,\rho(\tau),
\eeq
falling off as
\be
p(t)\stackunder{\approx}{t\to\infty}\frac{4c}{t^{1/4}}.
\ee
Thus, in Laplace space,
\beq\label{eq:sasy}
\lap{t}p(t)=\hat p(s)\stackunder{\approx}{s\to0}\frac{a}{s^{3/4}},
\eeq
with
\beq\label{eq:resa}
a=4\,\Gamma(3/4)\,c=\frac{3\,\eps}{2^{1/4}\sqrt{\pi}},
\eeq
from which the scaling form of $\hat\rho(s)$ follows
\beq
1-\hat\rho(s)=s\,\hat p(s)\approx as^{1/4}.
\label{scarho1}
\eeq

As a matter of interest, let us remark that
the expression of the survival probability (\ref{eq:survie}) has an explicit integral representation in Laplace space.
It reads (see, e.g.,~\cite{lachal,smedt})
\be
\hat p(s)
=\frac{1}{s}\bigac{1-\frac{3}{\pi}
\int_0^\infty\dd u\,\cosh u\frac{\sinh(3u/2)}{\sinh 3u}\cos\Big(v_0\sqrt{8s}\,\sinh u\Big)}.
\ee
For $s\to0$ this expression yields back (\ref{eq:sasy}), (\ref{eq:resa})~\cite{smedt}.

\section{Renewal structure and observables of interest}
\label{sec:renewal}

\subsection{Renewal structure for records}
\label{sec:renew}

A schematic drawing of a path of the process is depicted in figure \ref{fig:renew}.
This path is the continuum analogue of the path depicted in figure \ref{fig:figure0-X}.
Two series of conspicuous points are represented by red and green dots.
By convention, the origin is marked as a red dot.
The first green dot corresponds to the first crossing of the origin.
The following red dot is the first maximum of the path beyond this first green dot.
Then, considering the latter as the new origin, the process starts afresh.
In other words, sections between two red dots are probabilistic replicas of each other.
\textit{Record runs}, that is sequences of consecutive records of the discrete process,
translate into parts of these sections comprised between green and red dots.
This renewal structure will be the basis of all further considerations.
The red dots are renewal events (or renewals for short).

This can be formalised as follows (see figure \ref{fig:figure-def}).
The randomly accelerated particle is launched from the origin with a negative initial velocity $v_0=-\eps^2$,
where $\eps$ is a microscopic cutoff (see (\ref{epsdef}) and section \ref{sec:cutoff}).
The first section is made of the following sequence of points $(v_t,x_t)$:
\beq\label{eq:sequence}
(v_0,x_0=0)
\to\,
\stackunder{\underbrace{(\vone ,x_{\tone}=0)}}{\mathrm{green}}\,
\to\,
\stackunder{\underbrace{(v_{\tone +\mathbb{T}(\vone)}=0,x_{\tone +\mathbb{T}(\vone )}})}{\mathrm{red}},
\eeq
where, as defined earlier, $\tone $ is the epoch of first passage by the origin of $x_t$ and
$\mathbb{T}(\vone )$ is the hitting time of the origin for Brownian motion starting at $\vone$.
This velocity is positive, and typically much larger than the microscopic initial velocity $v_0$.
The last two points of the above sequence correspond respectively to the first green dot and to the following red dot.
This section is then repeated, i.e., the process starts afresh,
the particle is launched with initial velocity $v_0=-\eps^2$ from the last red dot considered as the new origin.

This scheme justifies in retrospect our choice of a negative initial velocity $v_0=-\eps^2$ at $t=0$, as mentioned above.
Had we taken another prescription at $t=0$,
this would have only changed the distribution of the first time interval (between $t=0$ and the first red dot),
keeping the distribution of all subsequent couples $(\tau_n,\delta_n)$ unchanged.
Changing the (somewhat arbitrary) initial conditions of the continuum process
would therefore only induce additional corrections to scaling falling off as $1/N_t\sim t^{-1/4}$ in relative value.

We shall denote by $\tau_1,\tau_2,\dots$ the successive copies of $\tone$ 
and by $\delta_1,\delta_2,\dots$ the successive copies of $\mathbb{T}(\vone )$.
We can view the process as being in one of two states: \textit{on} or \textit{off}. Initially it is \textit{off}, and it remains so for a time $\tau_1$; it then goes \textit{on} and remains so for a time $\delta_1$; and so forth.
The sequence of $\tau_n$ gives the lengths of the no-record runs, or \textit{off} states, while the sequence of $\delta_n$ gives the lengths of the record runs, or \textit{on} states.
Green and red dots signal the switchover points of the process from one state to the other.

To summarise, the cornerstone of the analysis that follows is the renewal
structure of the sequence of iid couples of intervals of time $(\tau_n,\delta_n)$
whose common density $\rho(\tau,\delta)$ is given in (\ref{eq:brique}).
Finally, we shall denote the waiting times between two renewals (that is, the duration of the sections between two red dots) as
$\sigma_n=\tau_n+\delta_n$.

\subsection{Factorisation of the joint distribution $\rho(\tau,\delta)$ in the asymptotic regime}
\label{sec:joint}

The expression (\ref{eq:brique}) of the density $\rho(\tau,\delta)$ actually exhibits a stronger form of scaling,
besides the power laws derived above.
This expression is indeed a homogeneous function of its arguments $\tau$ and $\delta$.
In other words, $\tone$ and the dimensionless ratio
\be
Z=\frac{\mathbb{T}(\vone )}{\tone }
\ee
become asymptotically independent as $\tone$ gets larger and larger.
Their joint law reads
\beq\label{eq:ftauz}
f_{\tone ,Z}(\tau,z)\approx\rho(\tau)f_Z\Big(z=\frac{\delta}{\tau}\Big),
\eeq
with
\beq\label{fzres}
f_Z(z)=\frac{12\,\Gamma(3/4)^2}{\pi^{3/2}}\,\frac{z^{1/4}}{(1+4z)^{7/4}}.
\eeq
The latter distribution is normalised, as should be.

In Laplace space, the transform $\hat\rho(s,u)$
of the joint density $\rho(\tau,\delta)$ has a scaling form which can be derived as follows.
We have
\bea
1-\hat\rho(s,u)
&=&\int_0^\infty\dd\tau\int_0^\infty\dd\delta\,\rho(\tau,\delta)(1-\e^{-(s\tau+u\delta)})
\\
&\approx&\int_0^\infty\dd z f_Z(z)\int_0^\infty\dd\tau\,\rho(\tau)(1-\e^{-(s+uz)\tau})
\\
&\approx&\int_0^\infty\dd z f_Z(z)\,(s+uz)\int_0^\infty\dd\tau\,p(\tau)\e^{-(s+uz)\tau}
\\
&\approx&\int_0^\infty\dd z f_Z(z)\,(s+uz)\,\hat p(s+uz)
\\
&\approx&a\int_0^\infty\dd z f_Z(z)\,(s+uz)^{1/4}.
\eea
The second line is derived from (\ref{eq:ftauz}),
the third one by means of an integration by parts,
and the fifth one by substituting $s+uz$ for $s$ in (\ref{eq:sasy}).

Introducing the dimensionless ratio
\beq
\xi=\frac{u}{s},
\label{xidef}
\eeq
we finally obtain
\beq
1-\hat\rho(s,u)\approx a s^{1/4} h(\xi),
\label{scarho}
\eeq
with
\beq
h(\xi)=\int_0^\infty\dd z f_Z(z)(1+\xi z)^{1/4}=\bigmean{(1+\xi Z)^{1/4}}.
\label{hint}
\eeq
This function admits the closed-form expression (see (\ref{hderiv}))
\beq\label{eq:hxi}
h(\xi)=\frac{1+\sqrt\xi}{\sqrt{1+\half\sqrt\xi}}.
\eeq
It is an algebraic function of degree four, obeying the biquadratic equation (see (\ref{apph4}))
\beq
(\xi-4)h^4+8h^2-4(\xi-1)^2=0.
\label{h4}
\eeq

A first consequence of the above is that the common distribution $f(\sigma)$
of the total waiting times $\sigma_n=\tau_n+\delta_n$
is also heavy-tailed with tail index 1/4, and amplitude proportional to $\eps$.
Its Laplace transform indeed reads
\beq
\hat f(s)=\hat\rho(s,s),
\label{hfsigma}
\eeq
thus from (\ref{scarho}) we get
\beq
1-\hat\rho(s,s)\approx a h(1) s^{1/4},
\label{hssasy}
\eeq
and therefore
\beq
f(\sigma)\approx\frac{c\,h(1)}{\sigma^{5/4}},
\label{fsigma}
\eeq
with (see (\ref{eq:meanR}), (\ref{eq:rave})), 
\beq\label{eq:resh1}
h(1)=\bigmean{(1+Z)^{1/4}}
=\frac{2\sqrt{6}}{3}=\frac{1}{1-\rr_\infty}.
\eeq
The tail index $1/4$ of this distribution is the same as that of the distributions of the intervals $\tau_n$ and $\delta_n$, i.e., it
is the persistence exponent of the random acceleration process.
Since this index is less than unity, the first moment of $f(\sigma)$
is divergent,
hence the renewal process built upon the waiting times $\sigma_n$ does not equilibrate,
but rather keeps a sensitive memory of its initial state.
The same holds for the complete renewal process built upon the couples $(\tau_n,\delta_n)$.
As a consequence, a large class of observables (see, e.g., section \ref{sec:obs}) are scale invariant.

\subsection{On the role of the cutoff}
\label{sec:cutoff}

 Let us come back to the prescription
which consists in launching the random acceleration process from $x_0=0$ with an initial microscopic velocity $v_0=-\eps^2$
both initially and at every red dot,
where $\eps$ is the cutoff defined in (\ref{epsdef}).

The integrated random walk process
breaks the continuum scale invariance of the random acceleration process,
if only because it is defined at discrete integer times $n$.
In order to get meaningful predictions from the continuum theory,
one must therefore break scale invariance by introducing a microscopic scale,
be it either spatial, temporal, or both.
A minimal prescription consists in imposing an initial velocity $v_0$.
This is manifest in the expression (\ref{eq:jointe}) of the joint law of $\tau_1$ and $v_1$.
This law degenerates to $\delta(\tau)\delta(v)$ in the $v_0\to0$ limit.

This is also manifest in the expression (\ref{eq:hitting0}) of the density of the first hitting time
$\mathbb{T}(v)$ of the origin for the Brownian velocity.
Intuitively, if the particle was launched at the origin with zero velocity, it would cross the origin almost immediately after, hence the hitting time of the origin could not be finite.

In some sense, the cutoff $\eps$ makes the connection between the discrete and continuum formalisms.
It enters the tail parameters of all power-law distributions:
(\ref{eq:ftauv}), (\ref{eq:rhotau}), (\ref{eq:fdelta}), (\ref{eq:brique}), (\ref{fsigma}),
as well as all non-universal results, such as the expression (\ref{nasy})
of the mean number of records.
The value of the cutoff $\eps$ appearing in these non-universal observables
turns out to depend on the distribution of step lengths (see Table \ref{tab:aeps}).
In this respect, the situation is quite similar to that met in
a recent study of the statistics of records for planar random walks~\cite{usplanar}.
There, too, it turns out to be necessary to introduce a cutoff into the continuum theory,
whose numerical value is different, e.g., for lattice Polya walks and
for off-lattice Pearson walks with steps of unit length.

\subsection{Observables of interest}
\label{sec:obs}

The observables studied in the sequel are defined as follows.

\subsubsection*{Number of renewals and epochs of last events.}

The first and simplest observable of interest, denoted by $N_{t}$,
is the number of renewals
(red dots in figure \ref{fig:renew}, discarding the red dot at the origin), or record runs, which occurred between $0$ and $t$, that is
 the random variable for the largest $n$ for which $T_{n}\le t$,
where the epoch $T_n$ of the $n-$th renewal is 
\beq\label{eq:tndef}
T_n=\underbrace{(\tau_1+\delta_1)}_{\sigma_1}
+\underbrace{(\tau_2+\delta_2)}_{\sigma_2}
+\cdots
+\underbrace{(\tau_n+\delta_n)}_{\sigma_n},
\eeq
and $T_0=0$.
For instance, $N_t=3$ in figure \ref{fig:renew}.

The epoch of the last renewal before $t$, that is of the $N_{t}-$th renewal, therefore reads
\beq\label{eq:tNdef}
T_{N_t}=\underbrace{(\tau_1+\delta_1)}_{\sigma_1}
+\underbrace{(\tau_2+\delta_2)}_{\sigma_2}
+\cdots
+\underbrace{(\tau_{N_t}+\delta_{N_t})}_{\sigma_{N_t}}.
\eeq
While $T_n$ is the sum of a fixed number $n$ of random variables $\sigma_n$,
$T_{N_t}$ is the sum of a random number $N_t$ of such random variables.

A related observable, denoted by $\w T_{N_t}$, is the epoch of the last dot before $t$, regardless of its colour (green or red), that is, the epoch of the last 
change of state of the process from \textit{off} to \textit{on} or from \textit{on} to \textit{off}, or else the last endpoint of an interval $\tau_n$ or $\delta_n$.

\subsubsection*{Number of records.}
Within the continuum formalism,
the number $M_n$ of records up to time $n$ of the integrated random walk 
is represented by the sum $S_t$
of all the intervals of time $\delta_n$ spent between green and red dots for $n=1,\dots,N_t$,
possibly up to a correction for the last interval.
In other words, $S_t$ is the total duration of all (complete or incomplete) record runs, or equivalently the total duration spent by the process in the \textit{on} state.

\medskip
Two cases are to be considered when dealing with
the quantities $S_t$ and $\w T_{N_t}$.
Either $t$ falls in the interval $\tau_{N_t+1}$, i.e., outside a record run as in figure \ref{fig:renew} (the process is \textit{off}),
or it falls in the interval $\delta_{N_t+1}$, i.e., inside a record run (the process is \textit{on}).
The first case occurs with asymptotic probability $1-\rr_\infty$, the second case with asymptotic probability $\rr_\infty$.

\vskip 4pt
\noindent
(i) In the first case, time $t$ is between a red and a green dot, hence $T_{N_t}<t<T_{N_t}+\tau_{N_t+1}$.
We have
\beqa\label{eq:first}
S_t&=&\delta_1+\cdots+\delta_{N_t},
\nonumber\\
\w T_{N_t}&=&T_{N_t}.
\eeqa
(ii) In the second case, time $t$ is between a green and a red dot, hence
 $T_{N_t}+\tau_{N_t+1}<t<T_{N_t}+\tau_{N_t+1}+\delta_{N_t+1}$.
We have
\beqa\label{eq:second}
S_t&=&\delta_1+\cdots+\delta_{N_t}+t-T_{N_t}-\tau_{N_t+1}
\nonumber\\
&=&t-(\tau_1+\cdots+\tau_{N_t+1}),
\nonumber\\
\w T_{N_t}&=&T_{N_t}+\tau_{N_t+1}.
\eeqa

The distributions of these observables will be determined in
 the following sections \ref{sec:nren}, \ref{sec:nrec}, \ref{sec:nepoch}.

\section{Number of renewals}
\label{sec:nren}

Following the definition given in the previous section,
for a given time $t$, $N_t$ is the unique integer such that
$T_{N_t}\le t<T_{N_t+1}$, with the definition (\ref{eq:tNdef}).
Let
\beq
p_n(t)=\prob(N_t=n)=\prob(T_n\le t<T_{n+1})
\eeq
denote the probability that $N_t$ equals some integer $n$.
In Laplace space, we have
\beqa
\hat p_n(s)
=
\lap{t} p_n(t)
&=&\left\langle\int_{T_n}^{T_{n+1}}\dd t\,\e^{-st}\right\rangle
\nonumber\\
&=&
\left\langle\frac{1-\e^{-s(\tau_{n+1}+\delta_{n+1})}}{s}\,\e^{-sT_n}\right\rangle
\nonumber\\
&=&
\frac{1-\hat\rho(s,s)}{s}\,\hat\rho(s,s)^n,
\label{eq:nlap}
\eeqa
which is well normalised.
This expression involves the joint law $\rho(\tau,\delta)$ only through the combination
$\hat\rho(s,s)=\hat f(s)$ (see (\ref{hfsigma}))
in accordance with the fact that the time intervals between successive renewals are the
total waiting times $\sigma_n=\tau_n+\delta_n$.

Let us focus our attention on the mean number $\mean{N_t}$ of renewals between 0 and $t$.
Its Laplace transform reads
\beq
\lap{t}\mean{N_t}
=\sum_{n\ge0}\hat p_n(s)=\frac{\hat\rho(s,s)}{s(1-\hat\rho(s,s))}.
\label{navelap}
\eeq
Using (\ref{hssasy}), this reads
\be
\lap{t}\mean{N_t}\approx\frac{1}{ah(1)s^{5/4}},
\ee
therefore 
\beq
\mean{N_t}\approx A\,t^{1/4},
\label{nasy}
\eeq
with (see (\ref{eq:resa}), (\ref{eq:resh1}))
\beq
A=\frac{1}{ah(1)\Gamma(5/4)}
=\underbrace{\frac{2^{3/4}\sqrt\pi}{\Gamma(1/4)\sqrt3}}_{0.474\,685\dots}\;\frac{1}{\eps}.
\label{avseps}
\eeq
The mean number of renewals grows as a power law
whose exponent 1/4 is the tail index of the law of the waiting times $\sigma_n$.
The predicted amplitude $A$ is the ratio of a universal number by the cutoff $\eps$.
Measuring the mean number of renewals therefore gives access to the value of $\eps$,
which is expected to depend on microscopic details of the discrete process,
i.e., on the parent distribution $\Phi$ of step lengths.

\begin{figure}[!ht]
\begin{center}
\includegraphics[angle=0,width=.7\linewidth,clip=true]{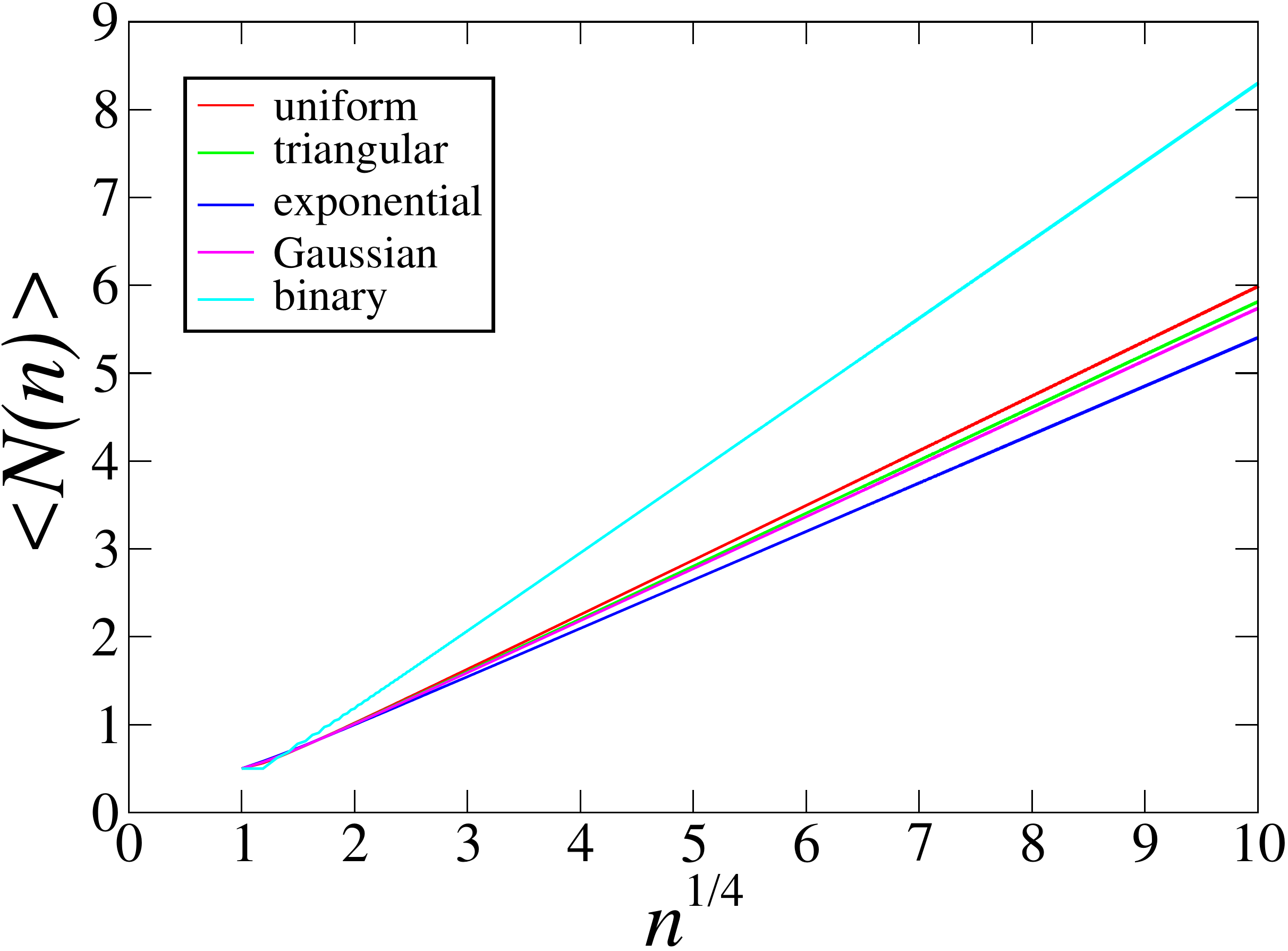}
\caption{\small
Mean number $\mean{N(n)}$ of renewals
of integrated random walks in discrete time against $n^{1/4}$ up to $n=10^4$,
for various symmetric distributions of the elementary steps (see legend).}
\label{nrenewals}
\end{center}
\end{figure}

Figure \ref{nrenewals} shows numerical data
for the mean number $\mean{N(n)}$ of renewals of integrated random walks
in discrete time against $n^{1/4}$,
for the step length distributions already considered in~Table \ref{q2}:
uniform, triangular, exponential, Gaussian and binary.
All datasets exhibit a very accurate asymptotic linear growth as a function of $n^{1/4}$.
The slopes $A$ of least-square fits over the range $10^2<n<10^4$ (regression lines are not shown)
and the corresponding values of the cutoff $\eps$,
according to (\ref{avseps}), are given in~Table \ref{tab:aeps}.
The binary distribution again appears as an outlier.

\begin{table}[!ht]
\begin{center}
\begin{tabular}{|l|c|c|}
\hline
distribution & $A$ & $\eps$ \cr
\hline
uniform & 0.623 & 0.762 \cr
triangular & 0.602 & 0.789 \cr
exponential & 0.551 & 0.861 \cr
Gaussian & 0.592 & 0.801 \cr
\hline
binary & 0.891 & 0.533 \cr
\hline
\end{tabular}
\caption
{Numerical values of the amplitude $A$ of the power-law growth (\ref{nasy})
of the mean number of renewals,
as extracted from the data shown in figure \ref{nrenewals},
and of the corresponding value of the cutoff $\eps$,
according to (\ref{avseps}),
for various symmetric step length distributions.}
\label{tab:aeps}
\end{center}
\end{table}

The full statistics of the number of renewals at large times
can be derived from an appropriate scaling analysis of the exact expression (\ref{eq:nlap}).
Omitting every detail, we obtain the following scaling formula~\cite{glrenew}
\be
N_t\approx\frac{1}{a h(1)}\,t^{1/4} X
\approx A\,\Gamma(5/4)\,t^{1/4} X,
\label{nx}
\ee
where the dimensionless reduced variable $X$ is distributed according to the universal law
\be
f_X(x)=\int\frac{\dd z}{2\pi\ii\,z^{3/4}}\,\e^{z-x z^{1/4}}.
\label{fcontour}
\ee
This probability density can be expressed
as a linear combination of three hyper\-geo\-metric functions
of type ${\null}_0F_2$~\cite{SM,Bar,Pen}.
We have the identity
\be
X\equiv(L_{1/4})^{-1/4},
\ee
where $L_{1/4}$ is distributed according to the one-sided L\'evy stable law of index $1/4$
and an appropriate scale factor.

\section{Number of records and total duration of record runs}
\label{sec:nrec}

The sum $S_t$ of the durations of all (complete or incomplete) record runs,
which is the continuum analogue of the number of records for integrated random walks, is the central observable of interest.
The purpose of this section is the determination of the density $ f_{S_t}(t,y)$ of $S_t$, defined as
\be
\prob\big(S_t\in(y,y+\dd y)\big)= f_{S_t}(t,y)\dd y.
 \ee
This quantity is entirely determined by the knowledge of the density $\rho(\tau,\delta)$.
In Laplace space we find that
\beqa\label{eq:laptxSt}
\hat f_{S_t}(s,u)&=&\lap {t}\lap {y}f_{S_t}(t,y)=\lap{t}\mean{\e^{-uS_t}}
\nonumber\\
&=&\frac{u(1-\hat\rho(s))+s(1-\hat\rho(s,s+u))}{s(s+u)(1-\hat\rho(s,s+u))},
\eeqa
as we now show by considering the two cases discussed in section \ref{sec:obs}.

\vskip4pt
\noindent
(i) In the first case, using (\ref{eq:first}),
\beqa\label{eq:i}
\hat f_{S_t}(s,u)_{(\rm i)}
&=&\sum_{n\ge0}\lap {t}\bigmean{\e^{-uS_t}I(T_n<t<T_n+\tau_{n+1})}
\nonumber\\
&=&
\sum_{n\ge0}\bigmean{\e^{-u(\delta_1+\cdots+\delta_n)}\int_{T_n}^{T_n+\tau_{n+1}}\dd t\,\e^{-st}}
\nonumber\\
&=&\sum_{n\ge0}\bigmean{\e^{-u(\delta_1+\cdots+\delta_n)}\e^{-sT_n}\frac{1-\e^{-s\tau_{n+1}}}{s}}
\nonumber\\
&=&\sum_{n\ge0}\bigmean{\e^{-s(\tau_1+\cdots+\tau_n)}\,
\e^{-(s+u)(\delta_1+\cdots+\delta_n)}\frac{1-\e^{-s\tau_{n+1}}}{s}
}
\nonumber\\
&=&\sum_{n\ge0}\hat\rho(s,s+u)^n\,\frac{1-\hat\rho(s)}{s}
\nonumber\\
&=&\frac{1}{1-\hat\rho(s,s+u)}\frac{1-\hat\rho(s)}{s}.
\eeqa
In the first line $I(\cdot)$ is the indicator function of the event in the parentheses.

\vskip4pt
\noindent
(ii) In the second case, using (\ref{eq:second}),
\beqa\label{eq:ii}
\hat f_{S_t}(s,u)_{(\rm ii)}
&=&\sum_{n\ge0}\lap {t}
\bigmean{\e^{-uS_t}I(T_n+\tau_{n+1}<t<T_n+\tau_{n+1}+\delta_{n+1})}
\nonumber\\
&=&\sum_{n\ge0}\bigmean{
\e^{u(\tau_1+\cdots+\tau_{n+1})}\int_{T_n+\tau_{n+1}}^{T_n+\tau_{n+1}+\delta_{n+1}}\dd t\,\e^{-(s+u)t}}
\nonumber\\
&=&\sum_{n\ge0}\hat\rho(s,s+u)^n\,\frac{\hat\rho(s)-\hat\rho(s,s+u)}{s+u}
\nonumber\\
&=&\frac{1}{1-\hat\rho(s,s+u)}\frac{\hat\rho(s)-\hat\rho(s,s+u)}{s+u}.
\eeqa
The two expressions (\ref{eq:i}) and (\ref{eq:ii}) add up to (\ref{eq:laptxSt}).

In order to analyse the sum $S_t$ at long times,
we consider the scaling regime where both Laplace variables $s$ and $u$ are small.
In this regime, the expression (\ref{eq:laptxSt})
can be simplified by means of the estimates (\ref{scarho1}) and (\ref{scarho}), i.e.,
\be
1-\hat\rho(s)\approx a s^{1/4},
\qquad {1-\hat\rho(s,s+u)}\approx a s^{1/4}h(\xi+1),
\ee
where $\xi=u/s$ and $h(\xi)$ is given in (\ref{eq:hxi}).
We thus obtain
\beq
\hat f_{S_t}(s,u)\approx\frac{1}{s}\,g_R(\xi),
\label{lapsca}
\eeq
with
\beqa
g_R(\xi)&=&\frac{1}{\xi+1}\left(1+\frac{\xi}{h(\xi+1)}\right)
\nonumber\\
&=&\frac{1}{\xi+1}\left(1+\left(1-\left(1-\half\xi\right)\sqrt{\xi+1}\right)^{1/2}\right).
\label{grres}
\eeqa
It can be shown by eliminating radicals
that $g_R$ is an algebraic function of degree four, obeying
\beq
4(\xi+1)^3g_R^4-16(\xi+1)^2g_R^3+16(\xi+1)g_R^2-(\xi-2)^2=0.
\label{gr4}
\eeq

The scaling expression (\ref{lapsca}) is entirely parameter-free.
It implies that, in the continuum theory, the asymptotic fraction of time spent during record runs (i.e., between green and red dots),
 \beq
R=\lim_{t\to\infty}\frac{S_t}{t},
\label{rdef}
\eeq
has a non-trivial universal distribution,
with density $f_R(x)$ defined by
\be
\prob\big(R\in(x,x+\dd x)\big)=f_R(x)\dd x\qquad (0<x<1).
\label{rlimit}
\ee
This random variable $R$ is also the limiting ratio
\beq\label{eq:RMnn}
R=\lim_{n\to\infty}\frac{M_n}{n},
\eeq
where $M_n$ is the total number of records of the integrated random walk up to discrete time $n$.
In particular, $\mean{R}=\rr_\infty$ (see (\ref{eq:Qinfty})). 

The existence of the limiting distribution $f_R(x)$ entails that
\bea
\hat f_{S_t}(s,u)&=&
\lap{t}\mean{\e^{-uS_t}}
\stackunder{\approx}{s,u\to0}\,\lap{t}\mean{\e^{-u tR}}
\\
&\approx&\lap{t}\int_0^\infty\dd x\,\e^{-u tx}f_R(x)
\approx\int_0^\infty\dd x\,f_R(x)\int_0^\infty\dd t\,\e^{-t(s+ux)}
\\
&\approx&\bigmean{\frac{1}{s+u R}}.
\eea
An identification with (\ref{lapsca}) yields
\beq
g_R(\xi)=\bigmean{\frac{1}{1+\xi R}}.
\label{grave}
\eeq

The explicit expression of $g_R(\xi)$ given in (\ref{grres})
allows the determination of the moments of $R$ as well as of its full distribution
(see \ref{appinvert} for details).

The moments of $R$ are readily derived by expanding $g_R(\xi)$ as a power series.
We thus obtain
\beq\label{eq:rave}
\mean{R}=1-\frac{1}{h(1)}=1-\frac{\sqrt{6}}{4}=0.387\,627\dots,
\eeq
and more generally
\beq
\mean{R^n}=1-a_n\sqrt{6},
\label{rmoms}
\eeq
where
\be
a_0=0,\quad a_1=\frac{1}{4},\quad
a_2=\frac{7}{24},\quad a_3=\frac{359}{1152},\quad
a_4=\frac{2239}{6912},\ \dots
\ee
These positive rational numbers obey the four-term linear recursion (see (\ref{aderiv}))
\beqa
(16n^2-9)a_{n-1}+(16n^2+48n+25)a_n
\nonumber\\
-16(n+1)(5n+7)a_{n+1}+48(n+1)(n+2)a_{n+2}=0.
\label{arec}
\eeqa

Using (\ref{eq:c}), one finds the following universal result
\beq
\label{eq:fR}
f_R(x)=\frac{\gamma_R(x)}{x^{3/4}(1-x)^{1/2}}\qquad(0<x<1)
\eeq
for the probability density of the asymptotic ratio $R$,
with
\beq
\gamma_R(x)=\frac{1+2x}{2\pi\sqrt{2x^{3/2}+\sqrt{1+3x}}}.
\label{gamrres}
\eeq
The limiting behaviours of the density $f_R(x)$ read
\be
f_R(x)\stackunder{\approx}{x\to0}\frac{1}{2\pi x^{3/4}},\quad
f_R(x)\stackunder{\approx}{x\to1}\frac{3}{4\pi(1-x)^{1/2}}.
\ee
The exponents appearing in the denominators
of these expressions can be interpreted as being equal to
$1-\th$, where $\th$ is a persistence exponent, by analogy with the singular
behaviour of the distribution of the occupation time of aging processes at its
two ends~\cite{drouffe1,drouffe2}.
For $x\to0$, $\th=1/4$, while for $x\to1$, $\th=1/2$.
The first case corresponds to the persistence exponent of the random acceleration process,
the second one to that of Brownian motion.

\begin{figure}[!ht]
\begin{center}
\includegraphics[angle=0,width=.7\linewidth,clip=true]{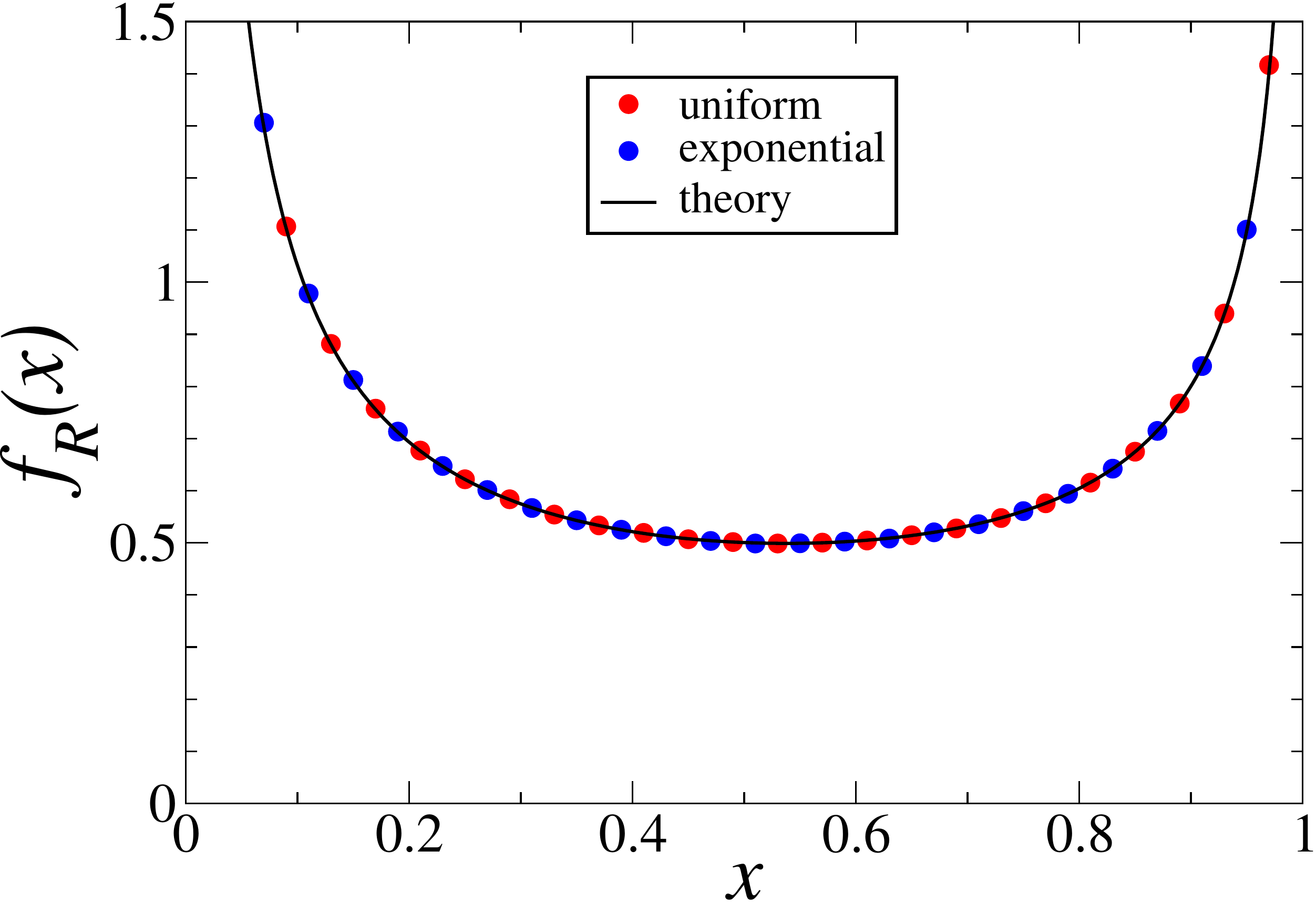}
\caption{\small
Symbols: histogram plots of the distribution of the total number of records
for integrated random walks made of $n=10^4$ steps,
with symmetric uniform and exponential step distributions (see legend),
rescaled according to (\ref{eq:RMnn}).
Full curve: theoretical prediction (\ref{eq:fR}).}
\label{fig:frhisto}
\end{center}
\end{figure}

Figure \ref{fig:frhisto} shows histogram plots of the distribution of the total number of records
for integrated random walks made of $n=10^4$ steps,
with symmetric uniform and exponential step distributions (see legend).
Numerical data are rescaled according to (\ref{eq:RMnn}).
Each dataset contains 50 bins.
Every second bin of each dataset is plotted alternatively.
Both rescaled histograms are in excellent agreement
with the theoretical prediction (\ref{eq:fR}) (full curve).

\section{Epochs of last events}
\label{sec:nepoch}
The last observables of interest are the two natural epochs
introduced in section \ref{sec:obs}.
The derivations of their distributions closely follow the approach described
in section \ref{sec:nrec}.

\subsection{Epoch $T_{N_t}$ of the last renewal}

The definition of the epoch $T_{N_t}$ of the last renewal (red dot) before time $t$
does not need to distinguish the two cases discussed in section \ref{sec:obs}.
The expression of the density in Laplace space is obtained in the manner set out above,
\bea
\hat f_{T_{N_t}}(s,u)
&=&
\sum_{n\ge0}\lap {t}\bigmean{\e^{-u T_{n}}I(T_n<t<T_{n+1})}
\nonumber\\
&=&
\sum_{n\ge0}\bigmean{\e^{-u T_n}\int_{T_n}^{T_{n+1}}\dd t\,\e^{-st}}
\nonumber\\
&=&
\sum_{n\ge0}\bigmean{\e^{-(s+u)T_n}
\frac{1-\e^{-s(\tau_{n+1}+\delta_{n+1})}}{s}}
\nonumber\\
&=&
\sum_{n\ge0}\bigmean{\e^{-(s+u)(\tau_1+\delta_1+\cdots+\tau_n+\delta_n)}
\frac{1-\e^{-s(\tau_{n+1}+\delta_{n+1})}}{s}}
\nonumber\\
&=&
\sum_{n\ge0}\hat\rho(s+u,s+u)^n\,\frac{1-\hat\rho(s,s)}{s}
\nonumber\\
&=&
\frac{1}{1-\hat\rho(s+u,s+u)}\frac{1-\hat\rho(s,s)}{s}.
\eea
This last expression involves the joint law $\rho(\tau,\delta)$ only through the combination
$\hat\rho(s,s)=\hat f(s)$ (see (\ref{hfsigma})).
This is again to be expected, as the time intervals between successive red dots are the
total waiting times $\sigma_n=\tau_n+\delta_n$.
Its asymptotic analysis proceeds as previously.
In the scaling regime where~$s$ and $u$ are small, we get
\be
\hat f_{T_{N_t}}(s,u)\approx\frac{1}{s}\,g_U(\xi),
\ee
where $\xi=u/s$ and
\beq
g_U(\xi)=(\xi+1)^{-1/4}.
\label{gures}
\eeq
The dimensionless ratio
\beq
U=\lim_{t\to\infty}\frac{T_{N_t}}{t}
\label{udef}
\eeq
therefore has a universal distribution,
such that
\be
g_U(\xi)=\bigmean{\frac{1}{1+\xi U}}.
\ee
The moments of $U$
\beq
\mean{U^n}=\frac{\Gamma(n+1/4)}{\Gamma(1/4)n!},
\label{umoms}
\eeq
are rational numbers, 
\be
\mean{U}=\frac{1}{4},\quad \mean{U^2}=\frac{5}{32},\quad
\mean{U^3}=\frac{15}{128},\quad \mean{U^4}=\frac{195}{2048},\ \dots.
\ee
The corresponding density reads
\beq
f_U(x)=\frac{1}{\pi\sqrt{2}}\,x^{-3/4}\,(1-x)^{-1/4}
=\beta_{\frac{1}{4},\frac{3}{4}}(x)\qquad(0<x<1),
\label{eq:fU}
\eeq
where
\be
\beta_{a,b}(x)=\frac{\Gamma (a+b)}{\Gamma (a)\Gamma (b)}x^{a-1}(1-x)^{b-1}
\ee
is the beta distribution on $[0,1]$.
We have thus recovered---in the present case for $\theta=1/4$---the beta distribution $f_U(x)=\beta_{1-\theta,\theta}(x)$
of the reduced epoch $U$ of the last renewal
for an arbitrary tail index $\theta<1$ (see, e.g.,~\cite{glrenew}).

\subsection{Epoch $\w T_{N_t}$ of the last change of state}

In Laplace space the expression of the density of $\w T_{N_t}$ reads
\beq\label{eq:lapv}
\hat f_{\w T_{N_t}}(s,u)
=\frac{1-\hat\rho(s)+\hat\rho(s+u)-\hat\rho(s+u,s)}
{s(1-\hat\rho(s+u,s+u))},
\eeq
as we now show by considering again the two cases discussed in section \ref{sec:obs}.

\vskip4pt
\noindent
(i) In the first case, using (\ref{eq:first}),
\beqa\label{eq:vi}
\hat f_{\w T_{N_t}}(s,u)_{(\rm i)}
&=&
\sum_{n\ge0}\lap {t}\bigmean{\e^{-uT_n}I(T_n<t<T_n+\tau_{n+1})}
\nonumber\\
&=&
\sum_{n\ge0}\bigmean{\e^{-uT_n}\int_{T_n}^{T_n+\tau_{n+1}}\dd t\,\e^{-st}}
\nonumber\\
&=&
\sum_{n\ge0}\bigmean{\e^{-(s+u)T_n}\frac{1-\e^{-s\tau_{n+1}}}{s}}
\nonumber\\
&=&
\sum_{n\ge0}\bigmean{\e^{-(s+u)(\tau_1+\delta_1+\cdots+\tau_n+\delta_n)}\,
\frac{1-\e^{-s\tau_{n+1}}}{s}}
\nonumber\\
&=&
\sum_{n\ge0}\hat\rho(s+u,s+u)^n\,\frac{1-\hat\rho(s)}{s}
\nonumber\\
&=&
\frac{1}{1-\hat\rho(s+u,s+u)}\frac{1-\hat\rho(s)}{s}.
\eeqa

\vskip4pt
\noindent
(ii) In the second case, using (\ref{eq:second}),
\beqa\label{eq:vii}
\hat f_{\w T_{N_t}}(s,u)_{(\rm ii)}
&=&
\sum_{n\ge0}\lap {t}
\bigmean{\e^{-u(T_n+\tau_{n+1})}I(T_n<t-\tau_{n+1}<T_n+\delta_{n+1})}
\nonumber\\
&=&
\sum_{n\ge0}
\bigmean{\e^{-u(T_n+\tau_{n+1})}\int_{T_n+\tau_{n+1}}^{T_n+\tau_{n+1}+\delta_{n+1}}\dd t\,\e^{-st}}
\nonumber\\
&=&
\sum_{n\ge0}\bigmean{\e^{-(s+u)(T_n+\tau_{n+1})}\frac{1-\e^{-s\delta_{n+1}}}{s}}
\nonumber\\
&=&
\sum_{n\ge0}\bigmean{\e^{-(s+u)(\tau_1+\delta_1+\cdots+\tau_n+\delta_n+\tau_{n+1})}\,
\frac{1-\e^{-s\delta_{n+1}}}{s}}
\nonumber\\
&=&
\sum_{n\ge0}\hat\rho(s+u,s+u)^n\,\frac{\hat\rho(s+u)-\hat\rho(s+u,s)}{s}
\nonumber\\
&=&
\frac{1}{1-\hat\rho(s+u,s+u)}\frac{\hat\rho(s+u)-\hat\rho(s+u,s)}{s}.
\eeqa
The two expressions (\ref{eq:vi}) and (\ref{eq:vii}) add up to (\ref{eq:lapv}).

The asymptotic analysis proceeds as previously.
In the scaling regime where~$s$ and $u$ are small, we get
\be
\hat f_{\w T_{N_t}}(s,u)\approx\frac{1}{s}\,g_V(\xi),
\ee
where $\xi=u/s$ and
\beqa
g_V(\xi)&=&\frac{1}{h(1)}
\left((\xi+1)^{-1/4}+h\left(\frac{1}{\xi+1}\right)-1\right)
\nonumber\\
&=&\frac{\sqrt{6}}{4}
\left((\xi+1)^{-1/4}+\frac{1+\sqrt{\xi+1}}{\sqrt{\xi+1+\half\sqrt{\xi+1}}}-1\right).
\label{gvres}
\eeqa
Note that the first term in the right side is equal to $(1-\rr_\infty)g_U(\xi)$, as it should.
The dimensionless ratio
\beq
V=\lim_{t\to\infty}\frac{\w T_{N_t}}{t}
\label{vdef}
\eeq
therefore has a universal distribution,
such that
\be
g_V(\xi)=\bigmean{\frac{1}{1+\xi V}}.
\ee

The moments of $V$ are readily derived by expanding $g_V(\xi)$ as a power series.
We thus obtain
\be
\mean{V}=\frac{\sqrt{6}}{16}+\frac{1}{6}=0.319\,759\dots,
\ee
and more generally
\beq
\mean{V^n}=(1-\rr_\infty)\mean{U^n}+b_n
\label{vmoms}
\eeq
for $n\ge1$ (see (\ref{umoms})), where
\be
b_1=\frac{1}{6},\quad b_2=\frac{11}{96},\quad
b_3=\frac{155}{1728},\quad b_4=\frac{12395}{165888},\ \dots
\ee
These positive rational numbers obey the four-term linear recursion (see (\ref{bderiv}))
\beqa
32(n-1)(2n-1)b_{n-1}-(176n^2+16n+5)b_n
\nonumber\\
+32(n+1)(5n+3)b_{n+1}-48(n+1)(n+2)b_{n+2}=0.
\label{brec}
\eeqa

Using again (\ref{eq:c}), some algebra yields the density
\beq
\label{eq:fV}
f_V(x)=\frac{\gamma_V(x)}{x^{3/4}(1-x)^{1/4}}\qquad(0<x<1),
\eeq
with
\beq
\gamma_V(x)=\frac{\sqrt3}{4\pi}
+\frac{\sqrt6}{4\pi x^{1/4}\sqrt{4-3x}}
\left(\sqrt{4-3x}-2(1-x)^{3/2}\right)^{1/2}.
\label{gamvres}
\eeq
Again one notes that the first contribution to $f_V$ is equal to $(1-\rr_\infty)f_U$.

\begin{figure}[!ht]
\begin{center}
\includegraphics[angle=0,width=.7\linewidth,clip=true]{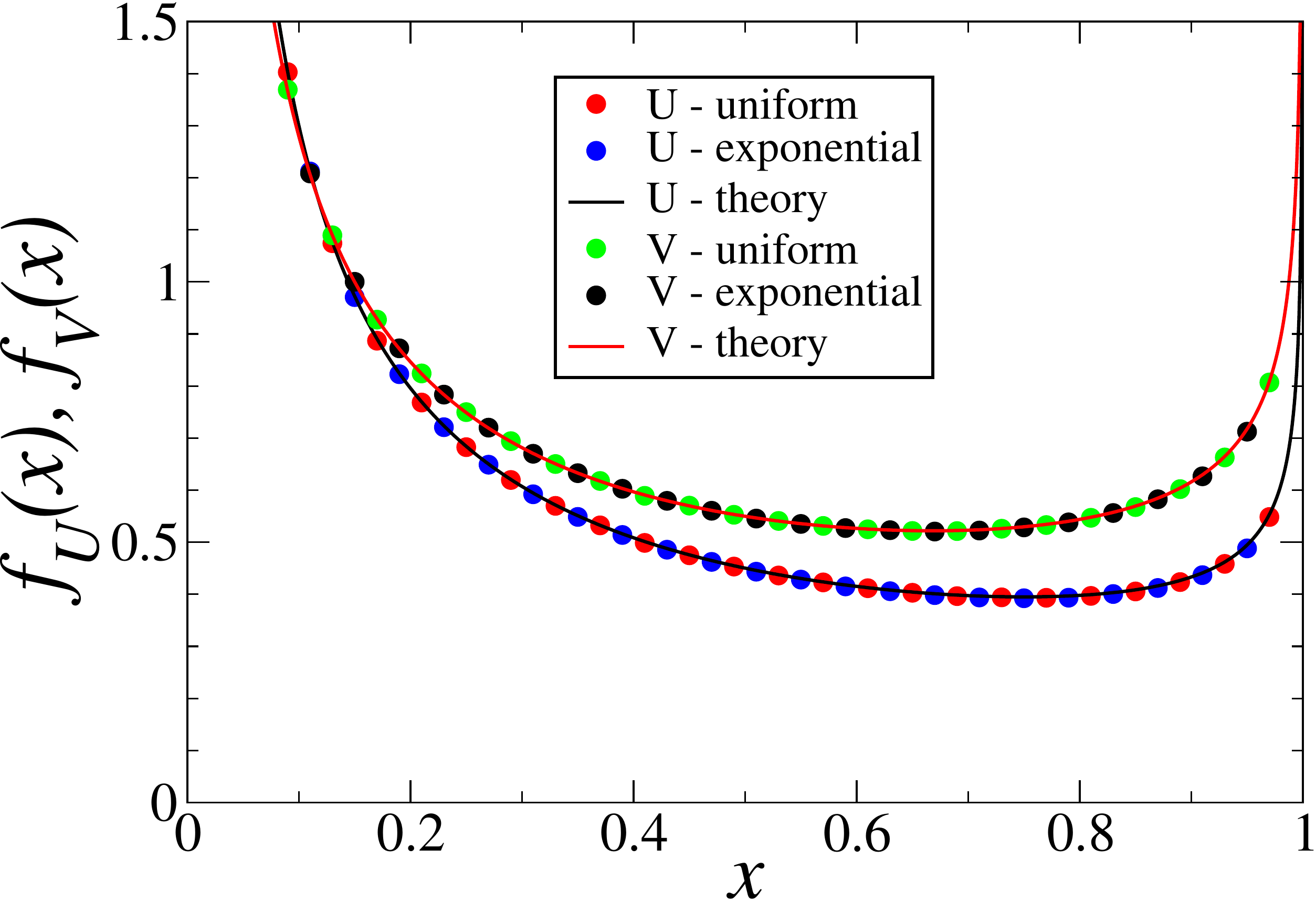}
\caption{\small
Symbols: histogram plots of the distribution of
the epochs $T_{N_t}$ of the last renewal and $\w T_{N_t}$ of the last marked point
for integrated random walks made of $n=10^4$ steps,
with symmetric uniform and exponential step distributions (see legend),
rescaled according to (\ref{udef}), (\ref{vdef}).
Full curves: predictions (\ref{eq:fU}), (\ref{eq:fV}).
}
\label{fig:fehisto}
\end{center}
\end{figure}

Figure \ref{fig:fehisto} shows histogram plots of the distribution of
the epochs $T_{N_t}$ and $\w T_{N_t}$ 
for integrated random walks made of $n=10^4$ steps,
with symmetric uniform and exponential step distributions (see legend).
Numerical data are rescaled according to (\ref{udef}), (\ref{vdef}).
Both rescaled histograms are again in excellent agreement
with the predictions (\ref{eq:fU}), (\ref{eq:fV}) (full curves).

\subsection{Time $t_{\rm m}$ to reach the maximum}

Let us denote by $t_{\rm m}$ the time at which the random acceleration process reaches its maximum 
during the time interval $(0,t)$,
and by
\be
W=\lim_{t\to\infty}\frac{t_{\rm m}}{t}
\ee
the rescaled time.
According to the dichotomy defined in section \ref{sec:obs}, either $t$ falls in the interval $\tau_{N_t+1}$,
or it falls in the interval $\delta_{N_t+1}$, with respective probabilities $1-\rr_\infty$ and $\rr_\infty$.
In the first case
$t_{\rm m}=T_{N_t}$, while in the second case $t_{\rm m}=t$.
It follows that
\beq\label{eq:tm}
f_W(x)=\rr_\infty\,\delta(x-1)+(1-\rr_\infty )f_U(x).
\eeq
We thus swiftly recover one of the main results of~\cite{rosso}, by a method which is completely different from that presented in this reference and which has the advantage of
 highlighting the probabilistic content of (\ref{eq:tm}).

\section{Discussion}
\label{discussion}

In this paper we have investigated the statistics of upper records
for integrated random walks with finite variance.
Our main focus was on the asymptotic regime of long times,
where the discrete integrated random walk process is described
by its continuum analogue, the random acceleration process.
Within this setting,
the asymptotic statistics of records 
relies on the renewal structure of the process,
involving a sequence of iid couples of intervals of time $(\tau_n,\delta_n)$,
defining an infinite alternating sequence of green and red dots.
Runs of consecutive records of the discrete process
take place in all time intervals $\delta_n$ between green and red dots.
For reference, let us mention that this alternating scheme is precisely what is referred to in the mathematical literature as an \textit{alternating renewal process}~\cite{cox}.
The alternating renewal process considered in the present work is generic in the sense that the \textit{off} and \textit{on} random variables, namely the time intervals $\tau_n$ and $\delta_n$, respectively, are not independent.

Let us come back on the three cases defined in the Introduction, namely iid random variables 
$\eta_1,\eta_2,\dots,\eta_n$,
the successive positions of a random walk
$V_1,V_2,\dots,V_n$, built from the previous variables, and the integrated random walk $X_1,X_2,\dots,X_n$, built from the latter.
The probability of record breaking at time $n$ and the mean number of records up to $n$ for these three cases are respectively
\beqa\label{eq:recap}
\rr_n= \frac{1}{n},&\qquad \mean{M_n} \stackunder{\approx}{n\to\infty} \ln n,
\nonumber\\
\rr_n\stackunder{\approx}{n\to\infty} \frac{1}{\sqrt{\pi\, n}},&\qquad \mean{M_n} \stackunder{\approx}{n\to\infty} \frac{2\sqrt{n}}{\sqrt{\pi}},
\nonumber\\
\rr_n\stackunder{\approx}{n\to\infty} \rr_\infty,&\qquad \mean{M_n} \stackunder{\approx}{n\to\infty} n\,\rr_\infty.
\eeqa
The classes of universality of these three cases are different and follow a hierarchy of decreasing generality.
For iid random variables, the parent distribution of step lengths $\Phi$ is assumed to be continuous.
For random walks, it is continuous and symmetric, thus
L\'evy flights belong to this class.
For integrated random walks, it is symmetric with finite variance.
For all three cases, discrete distributions have to be considered separately.
For integrated random walks universality only holds asymptotically, while for the other two cases it holds at any finite time.

In view of these considerations two natural extensions arise.
The first one consists in considering integrated random walks with heavy-tailed parent distribution $\Phi$.
The second one consists in considering $(k-1)$-fold integrated random walks.
For both cases, the question is to know how the last line of (\ref{eq:recap}) is modified.
We discuss both situations in turn.

 We first address the case of integrated L\'evy flights 
where the parent step length distribution is still symmetric, but
is now heavy-tailed, with a tail index $\alpha<2$,
so that $\mean{\eta^2}$ is divergent.
In this situation, the velocity and position of the walker scale respectively
as $V_n\sim n^{1/\alpha}$ and $X_n\sim n^{(\alpha+1)/\alpha}$.
The survival probability (\ref{eq:survie}),
which falls off as $t^{-1/4}$ in the diffusive case,
is known to decay as $t^{-\theta}$,
where the persistence exponent $\theta=1/(2+\alpha)$
varies continuously with $\alpha$~\cite{profeta}.
This situation is qualitatively different from the situation studied in this work.
Indeed, trajectories of integrated L\'evy flights are discontinuous in the continuum limit,
whereas those of the random acceleration process are continuous.
In spite of this, we expect that the number of records of integrated L\'evy flights
with $\alpha<2$ still grows linearly,
and that the corresponding ratio $R$ has a universal, $\alpha$-dependent, distribution.
In particular, the asymptotic probability of record breaking $\rr_\infty=\mean{R}$
is expected to be universal,
and to exhibit a smooth dependence on the tail index $\alpha$ over a rather small range,
interpolating between $\rr_\infty=0.387\,627\dots$ (see (\ref{eq:meanR})) for $\alpha=2$
and $\rr_\infty\to1/2$ as $\alpha\to0$.
In this limit, the process indeed grows so fast that the record-breaking condition at time $n$
becomes local and amounts to $V_n>0$.

It turns out that the case of $(k-1)$-fold integrated random walks
is somewhat similar.
Its continuum limit obeys $\dd^kx_t/\dd t^k=\eta_t$.
Usual random walks and integrated random walks considered in this work
respectively correspond to $k=1$ and $k=2$.
For general $k$, the position of the walker scales as $X_n\sim n^{k-1/2}$.
The survival probability decays as $t^{-\theta_k}$,
where the persistence exponent $\theta_k$ is a decreasing function of the integer~$k$.
Besides the exact values $\theta_1=1/2$ (Brownian motion) and
$\theta_2=1/4$ (random acceleration process),
we have $\theta_3\approx0.220$, $\theta_4\approx0.210$, $\theta_5\approx0.204$,
and so on~\cite{schwarz}.
The limit of these exponents, $\lim_{k\to\infty}\theta_k=3/16=0.1875$,
is related to the diffusion equation in the plane.
Its exact value has been derived only recently~\cite{popschehr,dornic}.
Here, too, the number of records is expected to grow linearly for all $k\ge2$,
with a universal, $k$-dependent, asymptotic probability of record breaking 
growing from $\rr_\infty=0.387\,627\dots$ for $k=2$
to $\rr_\infty\to1/2$ as $k\to\infty$,
for the same reason as above.

\ack
We acknowledge useful correspondence with A Lachal.

\appendix

\section{A word on notations}
\label{app:word}

\noindent \textit{Asymptotic equivalence}

The symbol $\approx$ stands for asymptotic equivalence;
the symbol $\sim$ is weaker and means `of the order of'.

\medskip
\noindent \textit{Probability densities, Laplace transforms, limiting distributions}

The probability density function of the continuous random variable $X$ is denoted by $f_X(x)$, with
\be
f_X(x)=\frac{\dd}{\dd x}\prob(X<x).
\ee
In the course of this work, we encounter several 
positive time-dependent continuous random variables,
denoted generically by $Y_{t}$.
The probability density function of such a random variable is denoted by $f_{Y_t}(t,y)$
where time $t$ appears as a parameter. 
The Laplace transform with respect to $y$ of 
this density is 
\be
\hat{f}_{Y_t}(t,u)=\lap{y}f_{Y}(t,y)
=\left\langle {\e}^{-uY_{t}}\right\rangle =\int_{0}^{\infty }\dd y\,{\e}^{-uy}\,f_{Y_t}(t,y), 
\ee
and its double Laplace transform with respect to $t$ and $y$
is denoted by 
\beq\label{app:double}
\fl
\hat{f}_{Y_t}(s,u)=\lap{t,y}f_{Y_t}(t,y)=\lap{t}\left\langle {\e}^{-uY_{t}}\right\rangle
=\int_{0}^{\infty }\dd t\,{\e}^{-st}\int_{0}^{\infty }\dd y\,{\e}^{-uy}\,f_{Y_t}(t,y).
\eeq
Assume that $Y_{t}$ scales 
asymptotically as $t$.
As $t\to\infty$ the density $f_{t^{-1}Y_t}(t,x=y/t)$ of the rescaled variable $Y_{t}/t$
 converges to a limit, denoted by 
\begin{equation}\label{eq:a}
f_{X}(x)=\lim_{t\rightarrow\infty}f_{t^{-1}Y_t}(t,x=y/t).
\end{equation}

\section{Inversion of the scaling form of a double Laplace transform}
\label{appinvert}

For completeness, we reproduce hereafter Appendix B of~\cite{glrenew}.

\vskip4pt\noindent
Consider the probability density function $f_{Y_t}(t,y)$ of the positive random
variable $Y_{t}$, and assume that its double Laplace transform (\ref{app:double}) with respect
to $t$ and $y$ has the scaling behaviour
\begin{equation}\label{hypothese}
\hat{f}_{Y_t}(s,u)=\frac{1}{s}\,
g\!\left(\frac{u}{s}\right)
\end{equation}
in the regime $s,u\rightarrow 0$, with $u/s$ arbitrary.
Then the following
properties hold.

\noindent (i) When $t\rightarrow\infty $ the random variable $Y_{t}/t$ possesses a limiting
distribution given by (\ref{eq:a}).

\noindent (ii) The scaling function $g$ is related to $f_{X}$ by
\begin{equation}\label{eq:b}
g(\xi)=\left\langle\frac{1}{1+\xi X}\right\rangle
=\int_{0}^{\infty
}\dd x\,\frac{f_{X}(x)}{1+\xi x}.
\end{equation}

\noindent (iii) This can be inverted as
\begin{equation}\label{eq:c}
f_{X}(x)=-\frac{1}{\pi x}\lim_{\epsilon\rightarrow 0}{\rm Im}\;
g\left(-\frac{1}{x+{\ii}\epsilon}\right).
\end{equation}
(iv) Finally the moments of $X$ can be obtained, when they exist, by expanding $g(\xi)$ as a
Taylor series, since (\ref{eq:b}) implies that
\begin{equation}\label{eq:d}
g(\xi)=\sum_{k\ge0}(-\xi)^{k}\left\langle X^{k}\right\rangle.
\end{equation}

These properties can be easily understood as follows.

\noindent (i)
First, a direct consequence of the scaling form (\ref{hypothese}) is that
$Y_{t}$ scales as $t$, as can be seen by Taylor expanding the right side of
this equation, which generates the moments of $Y_{t}$ in the Laplace space
conjugate to $t$.
Therefore (\ref{eq:a}) holds.

\noindent (ii)
Then, (\ref{eq:b}) is a simple consequence of (\ref{eq:a}), since
\be
\hat f_{Y_t}(s,u)=
\int_0^\infty\dd t\,\e^{-st}\mean{\e^{-uY_t}}
=\int_0^\infty\dd t\,\e^{-st}\mean{\e^{-u tX}}
=\bigmean{\frac{1}{s+uX}}.
\ee

\noindent (iii)
Now,
\be
f_{X}(x)=\bigmean{\delta\left(X-x\right)}
=-\frac{1}{\pi}\lim_{\epsilon\rightarrow 0}{\rm Im}\;\left\langle
\frac{1}{x+{\ii}\epsilon -X}\right\rangle.
\ee
The right side can be rewritten using (\ref{eq:b}), yielding (\ref{eq:c}).

\section{Some detailed derivations}
\label{appalgs}

This appendix is devoted to the detailed derivations of a few results used in the body of the paper.

\subsection{Derivation of the algebraic expression (\ref{eq:hxi}) of the function $h(\xi)$}
\label{appalgh}

The function $h(\xi)$ is defined by the integral expression (\ref{hint}),
where the distribution $f_Z(z)$ is given by (\ref{fzres}).
This reads
\be
h(\xi)=\frac{12\Gamma(3/4)^2}{\pi^{3/2}}\,
\int_0^\infty\dd z\,\frac{z^{1/4}(1+\xi z)^{1/4}}{(1+4z)^{7/4}}.
\ee
Setting $z=u/(4(1-u))$ and $\xi=4(1-\zeta)$, we obtain
\beqa
h&=&\frac{3\Gamma(3/4)^2}{\sqrt{2}\,\pi^{3/2}}\,
\int_0^1\dd u\,u^{1/4}(1-u)^{-3/4}(1-\zeta u)^{1/4}
\nonumber\\
&=&\frac{3}{\sqrt2}\,F\!\left(-\frac14,\frac54;\frac32;\zeta\right).
\label{hf}
\eeqa
The hypergeometric function boils down to something more elementary.
More precisely, we are facing the first of the 15 entries of the so-called Schwarz Table
of all cases where the hypergeometric series reduces
to an algebraic function (see, e.g.,~\cite[Vol.~I, Sec.~2.7.2]{htf}).

This reduction can be shown be elementary means as follows.
Starting from the hypergeometric differential equation obeyed by (\ref{hf}), i.e.,
\be
\zeta(1-\zeta)\frac{\dd^2h}{\dd\zeta^2}
+\left(\frac32-2\zeta\right)\frac{\dd h}{\dd\zeta}
+\frac{5}{16}h=0,
\ee
and setting $\xi=4\cos^2\alpha$, i.e., $\zeta=\sin^2\alpha$,
with $0\le\alpha\le\pi/2$ for definiteness, we obtain
\be
\frac{\dd^2h}{\dd\alpha^2}
+2\cot\alpha\,\frac{\dd h}{\dd\alpha}
+\frac{5}{16}\,h=0.
\ee
Setting $h=v/(\sin\alpha)$, we obtain the simple differential equation
\be
\frac{\dd^2v}{\dd\alpha^2}+\frac94\,v=0,
\ee
whose solutions are $\sin(3\alpha/2)$ and $\cos(3\alpha/2)$.
The regularity of $h$ and its value $h=1$ for $\xi=0$, i.e., $\alpha=\pi/2$, yield
\be
h=\sqrt2\,\frac{\sin(3\alpha/2)}{\sin\alpha}.
\ee
Some trigonometric identities finally yield
\beq
h=\frac{1+2\cos\alpha}{\sqrt{1+\cos\alpha}}
=\frac{1+\sqrt\xi}{\sqrt{1+\half\sqrt\xi}}.
\label{hderiv}
\eeq
It can be shown by eliminating radicals
that $h(\xi)$ is an algebraic function of degree four,
obeying the biquadratic equation
\beq
(\xi-4)h^4+8h^2-4(\xi-1)^2=0.
\label{apph4}
\eeq

\subsection{Derivation of the recursion (\ref{arec}) for the coefficients $a_n$}
\label{appalggr}

The gist of \ref{appalggr} and \ref{appalggv}
resides in the fact that algebraic functions
obey linear differential equations with polynomial coefficients.
As a consequence, the coefficients of their power-series expansions obey linear recursions.
These properties were known to Abel as early as 1827
(see~\cite{bostan} for an account of historical and algorithmic aspects).
In modern times they are only seldom mentioned or used.
The present case provides an example of a situation where they are useful.

The function $g_R(\xi)$ obeys the fourth-order algebraic equation (see (\ref{gr4}))
\beq
P(\xi,g_R)=4(\xi+1)^3g_R^4-16(\xi+1)^2g_R^3+16(\xi+1)g_R^2-(\xi-2)^2=0.
\label{appgr4}
\eeq
The linear differential equation obeyed by $g_R(\xi)$ can be derived in three steps.

First, its first derivative reads
\beqa
\frac{\dd g_R}{\dd\xi}
&=&-\frac{\partial P/\partial\xi}{\partial P/\partial g_R}
\nonumber\\
&=&-\frac{6(\xi+1)^2g_R^4-16(\xi+1)g_R^3+8g_R^2+2-\xi}{8(\xi+1)g_R((\xi+1)g_R-1)((\xi+1)g_R-2)}.
\label{gpex}
\eeqa
This expression is a rational function of $g_R$.
It can therefore be reduced to the form
\beq
\frac{\dd g_R}{\dd\xi}=A_3(\xi)g_R^3+A_2(\xi)g_R^2+A_1(\xi)g_R+A_0(\xi),
\label{gpan}
\eeq
where the $A_i$ are rational functions of $\xi$.
This can be done by expressing that the difference between (\ref{gpex}) and (\ref{gpan})
is a multiple of $P(\xi,g_R)$.
This condition yields coupled linear equations for the $A_i(\xi)$, whose solution yields
\beq
\frac{\dd g_R}{\dd\xi}=-\frac{N(\xi,g_R)}{4\xi(\xi+1)(\xi-2)(\xi-3)},
\label{gprime}
\eeq
with
\bea
N(\xi,g_R)
&=&12(\xi+1)^2g_R^3-36(\xi+1)g_R^2
\\
&+&(\xi^3-11\xi^2+243\xi+12)g_R+3(\xi-2)^2.
\eea

Second, higher-order derivatives of the function $g_R(\xi)$ can be readily evaluated
by applying iteratively the total derivative operator
\be
\frac{\dd}{\dd\xi}=\frac{\partial}{\partial\xi}+\frac{\dd g_R}{\dd\xi}\frac{\partial}{\partial g_R}
\ee
to the expression (\ref{gprime}).
In the present situation, it is sufficient to go up to the second derivative.

Third, eliminating nonlinear terms (those proportional to $g_R^2$ and $g_R^3$)
between the resulting expressions of the first and second derivatives,
we obtain the desired linear differential equation in the form
\beqa
16(\xi+1)^2(\xi-3)\frac{\dd^2g_R}{\dd\xi^2}
\nonumber\\
+16(\xi+1)(3\xi-7)\frac{\dd g_R}{\dd\xi}+(7\xi-25)g_R+9=0.
\label{grlin}
\eeqa

Finally, inserting the power-series expansion
\be
g_R(\xi)=\sum_{n\ge0}(1-a_n\sqrt6)(-\xi)^n
\ee
(see (\ref{grave}), (\ref{rmoms})) into (\ref{grlin}),
we obtain the following four-term linear recursion for the coefficients $a_n$:
\beqa
(16n^2-9)a_{n-1}+(16n^2+48n+25)a_n
\nonumber\\
-16(n+1)(5n+7)a_{n+1}+48(n+1)(n+2)a_{n+2}=0.
\label{aderiv}
\eeqa

\subsection{Derivation of the recursion (\ref{brec}) for the coefficients $b_n$}
\label{appalggv}

The following analysis is in the same vein as the previous section.
We start by splitting~$g_V(\xi)$ given in (\ref{gvres}) according to
\beq
g_V(\xi)=\frac{\sqrt{6}}{4}\left(g_U(\xi)-1\right)+g_B(\xi),
\label{gsplit}
\eeq
with
\be
g_B(\xi)=\frac{\sqrt{6}}{4}\frac{1+\sqrt{\xi+1}}{\sqrt{\xi+1+\half\sqrt{\xi+1}}}.
\ee
It can be shown by eliminating radicals
that $g_B$ is an algebraic function of degree four, obeying the biquadratic equation
\beq
16(\xi+1)(4\xi+3)g_B^4-48(\xi+1)^2g_B^2+9\xi^2=0.
\label{gb4}
\eeq

The linear differential equation obeyed by $g_B(\xi)$ can be derived
by means of the three-step procedure presented in \ref{appalggr}.
We thus obtain
\beq
16(\xi+1)^2(4\xi+3)\frac{\dd^2g_B}{\dd\xi^2}+96(\xi+1)^2\frac{\dd g_B}{\dd\xi}+5g_B=0.
\label{gblin}
\eeq

Inserting the power-series expansion
\be
g_B(\xi)=\sum_{n\ge0}b_n(-\xi)^n
\ee
(see (\ref{vmoms}), (\ref{gsplit})) into (\ref{gblin}),
we obtain the following four-term linear recursion for the coefficients $b_n$:
\beqa
32(n-1)(2n-1)b_{n-1}-(176n^2+16n+5)b_n
\nonumber\\
+32(n+1)(5n+3)b_{n+1}-48(n+1)(n+2)b_{n+2}=0.
\label{bderiv}
\eeqa
We have $b_0=1$,
whereas the $b_n$ enter the expression (\ref{vmoms}) of the moments of $V$ for $n\ge1$ only.

\section*{References}

\bibliography{paper.bib}

\providecommand{\newblock}{}
\begin{thebibliography}{10}
\expandafter\ifx\csname url\endcsname\relax
  \def\url#1{{\tt #1}}\fi
\expandafter\ifx\csname urlprefix\endcsname\relax\def\urlprefix{URL }\fi
\providecommand{\eprint}[2][]{\url{#2}}

\bibitem{chandler}
Chandler K~N 1952 {\em J. Roy. Statist. Soc.: Series B\/} {\bf 14} 220--228

\bibitem{renyi}
R\'enyi A 1962 {\em Ann. Sci. Univ. Clermont-Ferrand\/} {\bf 8} 7--13

\bibitem{foster}
Foster F~G and Stuart A 1954 {\em J. Roy. Statist. Soc.: Series B\/} {\bf 16}
  1--13

\bibitem{glick}
Glick N 1978 {\em Amer. Math. Monthly\/} {\bf 85} 2--26

\bibitem{arnold}
Arnold B~C, Balakrishnan N and Nagaraja H~N 1998 {\em Records\/} (New York:
  Wiley)

\bibitem{nevzorov2}
Nevzorov V~B 2001 {\em Records: Mathematical Theory (Translation of
  Mathematical Monographs vol {\bf 194})\/} (Providence, RI: American
  Mathematical Society)

\bibitem{bunge}
Bunge J and Goldie C~M 2001 {\em Handbook of Statistics\/} {\bf 19} 277--308

\bibitem{nevzorov}
Nevzorov V~B and Balakrishnan N 1998 {\em Handbook of Statistics\/} {\bf 16}
  515--570

\bibitem{blackwell}
Blackwell D 1953 {\em Pacific J. Math.\/} {\bf 3} 315--320

\bibitem{feller1}
Feller W 1957 {\em An Introduction to Probability Theory and its
  Applications\/} 2nd ed vol~1 (New York: Wiley)

\bibitem{spitzer}
Spitzer F 2001 {\em Principles of Random Walk\/} (New York: Springer)

\bibitem{feller2}
Feller W 1971 {\em An Introduction to Probability Theory and its
  Applications\/} 2nd ed vol~2 (New York: Wiley)

\bibitem{cox}
Cox D~R 1962 {\em Renewal Theory\/} (London: Methuen)

\bibitem{cox-miller}
Cox D~R and Miller H~D 1965 {\em The Theory of Stochastic Processes\/} (London:
  Chapman \& Hall)

\bibitem{sparre53}
Sparre~Andersen E 1953 {\em Math. Scand.\/} {\bf 1} 263--285

\bibitem{sparre54}
Sparre~Andersen E 1954 {\em Math. Scand.\/} {\bf 2} 194--222

\bibitem{glrenew}
Godr\`eche C and Luck J~M 2001 {\em J. Stat. Phys.\/} {\bf 104} 489--524

\bibitem{ziff}
Majumdar S~N and Ziff R~M 2008 {\em Phys. Rev. Lett.\/} {\bf 101} 050601

\bibitem{revue}
Godr\`eche C, Majumdar S~N and Schehr G 2017 {\em J. Phys. A: Math. Theor.\/}
  {\bf 50} 333001

\bibitem{wergen}
Wergen G 2013 {\em J. Phys. A: Math. Theor.\/} {\bf 46} 223001

\bibitem{mckean}
McKean H~P 1962 {\em Kyoto J. Math.\/} {\bf 2} 227--235

\bibitem{goldman}
Goldman M 1971 {\em Ann. Math. Statist.\/} {\bf 42} 2150--2155

\bibitem{marshall}
Marshall T and Watson E 1985 {\em J. Phys. A: Math. Theor.\/} {\bf 18} 3531

\bibitem{lachal}
Lachal A 1991 {\em Annales de l'IHP Probabilit{\'e}s et statistiques\/} {\bf
  27} 385--405

\bibitem{sinai}
Sinai Y~G 1992 {\em Theor. Math. Phys.\/} {\bf 90} 219--241

\bibitem{profeta}
Profeta C and Simon T 2015 {\em Probab. Theory Relat. Fields\/} {\bf 162}
  463--485

\bibitem{burk1}
Burkhardt T~W 1993 {\em J. Phys. A: Math. Theor.\/} {\bf 26} L1157--L1162

\bibitem{bray}
Swift M~R and Bray A~J 1999 {\em Phys. Rev. E\/} {\bf 59} R4721--R4724

\bibitem{burk2}
Burkhardt T~W 2000 {\em J. Phys. A: Math. Theor.\/} {\bf 33} L429--L432

\bibitem{smedt}
\protect{De Smedt} G, Godr\`eche C and Luck J~M 2001 {\em Europhys. Lett.\/}
  {\bf 53} 438--443

\bibitem{rosso}
Majumdar S~N, Rosso A and Zoia A 2010 {\em J. Phys. A: Math. Theor.\/} {\bf 43}
  115001

\bibitem{burk3}
{Burkhardt} T~W 2017 {\em J. Stat. Phys.\/} {\bf 169} 730--743

\bibitem{singh}
Singh P 2020 {\em J. Phys. A: Math. Theor.\/} {\bf 53} 405005

\bibitem{burk4}
Burkhardt T~W 2017 First passage of a randomly accelerated particle {\em
  First-Passage Phenomena and Their Applications\/} ed Metzler R, Oshanin G and
  Redner S (Singapore: World Scientific) chap~2, pp 21--44 (\textit{Preprint}
  \eprint{arXiv:1603.07017})

\bibitem{usplanar}
Godr\`eche C and Luck J~M 2021 {\em J. Phys. A: Math. Theor.\/} {\bf 54} 325003

\bibitem{SM}
Scher H and Montroll E~W 1975 {\em Phys. Rev. B\/} {\bf 12} 2455--2477

\bibitem{Bar}
Barkai E 2001 {\em Phys. Rev. E\/} {\bf 63} 046118

\bibitem{Pen}
Penson K~A and G\'orska K 2010 {\em Phys. Rev. Lett.\/} {\bf 105} 210604

\bibitem{drouffe1}
Drouffe J~M and Godr{\`e}che C 1998 {\em J. Phys. A: Math. Theor.\/} {\bf 31}
  9801--9807

\bibitem{drouffe2}
Drouffe J~M and Godr{\`e}che C 2001 {\em Eur. Phys. J. B\/} {\bf 20} 281--288

\bibitem{schwarz}
Schwarz J~M and Maimon R 2001 {\em Phys. Rev. E\/} {\bf 64} 016120

\bibitem{popschehr}
Poplavskyi M and Schehr G 2018 {\em Phys. Rev. Lett.\/} {\bf 121} 150601

\bibitem{dornic}
Dornic I 2018 \protect{Universal Painlev\'e VI probability distribution in
  Pfaffian persistence and Gaussian first-passage problems with a sech-kernel}
  (\textit{Preprint} \eprint{arXiv:1810.06957})

\bibitem{htf}
Erd\'elyi A 1953 {\em Higher Transcendental Functions (The Bateman Manuscript
  Project)\/} (New York: McGraw-Hill)

\bibitem{bostan}
Bostan A, Chyzak F, Salvy B, Lecerf G and Schost E 2007 Differential equations
  for algebraic functions {\em Proceedings of the 2007 International Symposium
  on Symbolic and Algebraic Computation\/} (New York: Association for Computing
  Machinery) (\textit{Preprint} \eprint{arXiv:cs/0703121})

\end{thebibliography}

\end{document}